%% file: 2022_noma_cuda_demonstrator_arXiv.tex
\newcommand{\comment}[1]{}

\documentclass{article}
\usepackage{arxiv}

\usepackage[T1]{fontenc}
\usepackage[utf8]{inputenc}
\usepackage[english]{babel}

\input{settings/acronym}
\input{settings/bib}

\input{settings/math}

\usepackage{orcidlink}

\input{settings/title}
\input{settings/hyperref}
\input{settings/cleveref}

\input{settings/graphic}
\input{settings/color}
\input{settings/table}
\input{settings/tikz}

\input{settings/siunix}
\usepackage[super]{nth}
\usepackage{textcomp}

\usepackage{hyphenat}

\newtheorem{fact}{Fact}
\newtheorem{remark}{Remark}

\begin{document}


\comment{%
\input{sections/abstract.tex}

\titlepgskip=-7ex
\maketitle

}%


\comment{%
\maketitle

\nohyphens{ 
\input{sections/abstract.tex}
}

}%


\maketitle

\nohyphens{ 
\input{sections/abstract.tex}
}



\acresetall
\newpage

\begingroup
\let\clearpage\relax
\nohyphens{
\include{sections/I_introduction}
\include{sections/II_background}
\include{sections/III_implementation}
\include{sections/IV_lab_setup}
\include{sections/V_results}
\include{sections/VI_conclusions}
\include{sections/acknowledgments}
}
\include{sections/references}

\endgroup

\end{document}

%% file: settings/acronym.tex
\usepackage[printonlyused,nolist]{acronym}

\AtBeginDocument{

\begin{acronym}

    \acro{4G}[4G]{4th generation}
    \acro{5G}[5G]{5th generation}
    \acro{6G}[6G]{6th generation}
    \acro{5GPPP}[5G-PPP]{5G infrastructure public private partnership}
    \acro{3GPP}[3GPP]{3rd Generation Partnership Project}
    \acro{ADC}[ADC]{analog-to-digital converter}
    \acro{APSM}[APSM]{adaptive projected subgradient method}
    \acro{ASIC}[ASIC]{Application-Specific Integrated Circuit}
    \acro{BER}[BER]{bit error rate}
    \acro{BPSK}[BPSK]{binary phase-shift keying}
    \acro{BS}[BS]{base station}
    \acro{CFO}[CFO]{carrier frequency offset}
    \acro{COMP}[CoMP]{coordinated multi-point}
    \acro{COTS}[COTS]{commercial off-the-shelf}
    \acro{CPU}[CPU]{central processing unit}
    \acro{DAC}[DAC]{digital-to-analog converter}
    \acro{DL}[DL]{downlink}
    \acro{DMA}[DMA]{direct memory access}
    \acro{DU}[DU]{Distributed Unit}
    \acro{EC}[EC]{European commission}
    \acro{eNB}[eNB]{evolved Node B}
    \acro{FPGA}[FPGA]{field-programmable gate array}
    \acro{FR1}[FR1]{Frequency Range 1}
    \acro{FR2}[FR2]{Frequency Range 2}
    \acro{FEC}[FEC]{forward error correction} 
    \acro{gNB}[gNB]{next generation Node B}
    \acro{GPS}[GPS]{global positioning system}
    \acro{GPU}[GPU]{graphics processing unit}
    \acro{H2020}[H2020]{Horizon 2020}
    \acro{HPC}[HPC]{high-performance computing}
    \acro{IIO}[IIO]{industrial I/O}
    \acro{IOT}[IoT]{internet of things}
    \acro{LO}[LO]{local oscillator}
    \acro{LTE}[LTE]{long-term evolution}
    \acro{MAP}[MAP]{maximum a-posteriori}
    \acro{MCM}[MCM]{multi-carrier modulation}
    \acro{ML}[ML]{machine learning}
    \acro{MIMO}[MIMO]{multiple-input multiple-output}
    \acro{MMIMO}[mMIMO]{massive multiple-input multiple-output}
    \acro{MMSE}[MMSE]{minimum mean square error}
    \acro{MMTC}[mMTC]{massive machine-type communication}
    \acro{NOMA}[NOMA]{non-orthogonal multiple access}
    \acro{MU}[MU]{multiuser}
    \acro{MA}[MA]{multiple access}
    \acro{MUSA}[MUSA]{multi-user Shared Access}
    \acro{NI}[NI]{National Instruments}
    \acro{NL}[NL]{nonlinear}
    \acro{NR}[NR]{New Radio}
    \acro{NV}[NV]{Nvidia}
    \acro{OFDM}[OFDM]{orthogonal frequency-division multiplexing}
    \acro{OFDMA}[OFDMA]{orthogonal frequency-division multiple access}
    \acro{OMA}[OMA]{orthogonal multiple access}
    \acro{OTB-5G+}[OTB-5G+]{Open Testbed Berlin - 5G and Beyond}
    \acro{PC}[PC]{personal computer}
    \acro{PLL}[PLL]{phase lock loop}
    \acro{POCS}[POCS]{projection onto convex sets}
    \acro{PoC}[PoC]{proof of concept}
    \acro{QPSK}[QPSK]{quadrature phase-shift keying}
    \acro{RAN}[RAN]{Radio Access Network}
    \acro{RKHS}[RKHS]{reproducing kernel Hilbert space}
    \acro{RRH}[RRH]{remote radio head}
    \acro{Rx}[Rx]{receiver}
    \acro{RRC}[RRC]{root-raised-cosine}
    \acro{SCM}[SCM]{Single-Carrier Modulation}
    \acro{SC-FDMA}[SC-FDMA]{Single-Carrier Frequency-Division Multiple Access}
    \acro{SDR}[SDR]{Software-Defined Radio}
    \acro{SIC}[SIC]{Successive Interference Cancellation}
    \acro{SIMO}[SIMO]{Single-Input Multiple-Output}
    \acro{SIMT}[SIMT]{single instruction multiple threads}
    \acro{SINR}[SINR]{signal-to-interference-plus-noise ratio}
    \acro{SNR}[SNR]{signal-to-noise ratio}
    \acro{SoC}[SoC]{System-on-Chip}
    \acro{SER}[SER]{Symbol Error Rate}
    \acro{TS}[TS]{technical specification}
    \acro{Tx}[Tx]{transmitter}
    \acro{UE}[UE]{user equipment}
    \acro{UCA}[UCA]{Uniform Circular Array}
    \acro{UL}[UL]{uplink}
    \acro{ULL}[ULL]{ultra-low latency}
    \acro{URLLC}[URLLC]{ultra-reliable and low latency communications}
    \acro{USRP}[USRP]{Universal Software Radio Peripheral}
    \acro{UHD}[UHD]{USRP Hardware Driver}
    \acro{UPA}[UPA]{Uniform Planar Array}
    \acro{WLAN}[WLAN]{Wireless Local Area Network}
    \acro{CUDA}[CUDA]{Compute Unified Device Architecture}
    \acro{GPC}[GPC]{Graphics Processing Cluster}
    \acro{TPC}[TPC]{Texture Processing Cluster}
    \acro{SM}[SM]{Streaming Multiprocessor}
    \acro{RT}[RT]{ray tracing}
    \acro{TDP}[TDP]{Thermal Design Power}
    \acro{GDDR}[GDDR]{Graphics Double Data Rate}
    \acro{HBM}[HBM]{High Bandwidth Memory}
    \acro{MATLAB}[MATLAB]{MATrix LABoratory}
    
\end{acronym}

}

\usepackage{xspace}

\makeatletter
\DeclareRobustCommand\onedot{\futurelet\@let@token\@onedot}
\def\@onedot{\ifx\@let@token.\else.\null\fi\xspace}

\def\eg{e.\,g\onedot}

\def\ie{i.\,e\onedot} 
 
\def\etc{etc\onedot}

\makeatother

\newcommand{\signal}[1]{{\boldsymbol{#1}}}

\newcommand{\Natural}{{\mathbb N}}


%% file: settings/bib.tex
\usepackage{cite}

\usepackage{url}

\def\BibTeX{{\rm B\kern-.05em{\sc i\kern-.025em b}\kern-.08em
    T\kern-.1667em\lower.7ex\hbox{E}\kern-.125emX}}



%% file: settings/math.tex
\usepackage{amsmath,amssymb,amsfonts,amsthm}
\usepackage{algorithmic}

\usepackage{nicefrac}
\usepackage{mathtools}


%% file: settings/title.tex
\newcommand{\PAPERtitle}{GPU-accelerated partially linear multiuser detection for 5G and beyond URLLC systems}

\newcommand{\PAPERauthorOne}{Matthias Mehlhose}
\newcommand{\PAPERemailOne}{matthias.mehlhose}
\newcommand{\PAPERorcidOne}{0000-0002-8307-7120}

\newcommand{\PAPERauthorTwo}{Daniel Schäufele}
\newcommand{\PAPERemailTwo}{daniel.schaeufele}

\newcommand{\PAPERauthorThree}{Daniyal Amir Awan}
\newcommand{\PAPERemailThree}{daniyal.a.awan}

\newcommand{\PAPERauthorFour}{Guillermo Marcus}
\newcommand{\PAPERemailFour}{gmarcus}

\newcommand{\PAPERauthorFive}{Nikolaus Binder}
\newcommand{\PAPERemailFive}{nbinder}

\newcommand{\PAPERauthorSix}{Martin Kasparick}
\newcommand{\PAPERemailSix}{martin.kasparick}

\newcommand{\PAPERauthorSeven}{Renato L. G. Cavalcante}
\newcommand{\PAPERemailSeven}{renato.cavalcante}

\newcommand{\PAPERauthorEight}{S{\l}awomir Sta{\'n}czak}
\newcommand{\PAPERemailEight}{slawomir.stanczak}

\newcommand{\PAPERauthorNine}{Alexander Keller}
\newcommand{\PAPERemailNine}{akeller}

\newcommand{\PAPERemailHHI}{hhi.fraunhofer.de}
\newcommand{\PAPERemailTU}{tu-berlin.de}
\newcommand{\PAPERemailNV}{nvidia.com}

\newcommand\PAPERkeywords{Machine learning, Wireless communication, Multiuser detection, NOMA, MIMO, Ultra-reliable low latency communication, Massively parallel architectures, GPU, CUDA}

\makeatletter
\@ifclassloaded{ieeeaccess}
{%
    \AtBeginDocument{%
        \history{Date of publication mmmm dd, 2021,
                 date of current version \today.}
        \doi{10.1109/ACCESS.2021.DOI}
        \title{\PAPERtitle}
        \author{%
            \mbox{\PAPERauthorOne}\authorrefmark{1},
            \mbox{\PAPERauthorFour}\authorrefmark{3},
            \mbox{\PAPERauthorTwo}\authorrefmark{1},
            \mbox{\PAPERauthorThree}\authorrefmark{1},
            \mbox{\PAPERauthorFive}\authorrefmark{3},
            \mbox{\PAPERauthorSix}\authorrefmark{1},
            \mbox{\PAPERauthorSeven}\authorrefmark{1},
            \mbox{\PAPERauthorEight}\authorrefmark{1,2}
            and
            \mbox{\PAPERauthorNine}\authorrefmark{3}
        }
        \address[1]{Fraunhofer Institute for Telecommunications, Heinrich Hertz Institute (HHI),
        Department of Wireless Communications and Networks,\\ Einsteinufer 37, 10587 Berlin, Germany}
        \address[2]{Technical University Of Berlin, Faculty IV - Electrical Engineering and Computer Science,
        Department of Telecommunication Systems,\\ Einsteinufer 27, 10587 Berlin, Germany}
        \address[3]{NVIDIA,\\ Fasanenstrasse 81, 10623 Berlin, Germany}

        \tfootnote{This work has been partially funded by the German Federal Ministry of Education and Research (BMBF, Germany) in the project \ac{OTB-5G+} under Grant 16KIS0980
                   and supported as part of the 6G Research and Innovation Cluster 6G-RIC under Grant 16KISK020K.}

        \markboth
        {\PAPERauthorOne\space\headeretal: \PAPERtitle}
        {\PAPERauthorOne\space\headeretal: \PAPERtitle}
        \corresp{Corresponding author: \PAPERauthorOne\orcidlink{\PAPERorcidOne}\space 
        (e-mail: \PAPERemailOne{@}\PAPERemailHHI).}
        
    }
}
{
\@ifclassloaded{IEEEtran}
{%
    \AtBeginDocument{%
        \title{\PAPERtitle}
        \author{%
            \IEEEauthorblockN{\PAPERauthorOne\IEEEauthorrefmark{1}, \PAPERauthorFour\IEEEauthorrefmark{3},
            \PAPERauthorTwo\IEEEauthorrefmark{1}, \PAPERauthorThree\IEEEauthorrefmark{1}\IEEEauthorrefmark{2},  \PAPERauthorFive\IEEEauthorrefmark{3}}
            \IEEEauthorblockN{\PAPERauthorSix\IEEEauthorrefmark{1}, \PAPERauthorSeven\IEEEauthorrefmark{1}\IEEEauthorrefmark{2}, \PAPERauthorEight\IEEEauthorrefmark{1}\IEEEauthorrefmark{2} and \PAPERauthorNine\IEEEauthorrefmark{3}}
            \vspace{0.05in}
            \IEEEauthorblockA{\IEEEauthorrefmark{1}
                \textit{Fraunhofer Institute for Telecommunications},
                \textit{Heinrich Hertz Institute (HHI)}\\
                Department of Wireless Communications and Networks,
                Einsteinufer 37, 10587 Berlin, Germany\\
                \{\PAPERemailOne, \PAPERemailTwo, \PAPERemailSix, \PAPERemailSeven, \PAPERemailEight \}@\PAPERemailHHI
            }
            \vspace{0.05in}
            \IEEEauthorblockA{\IEEEauthorrefmark{2}
                \textit{Technical University Of Berlin},
                \textit{Faculty IV - Electrical Engineering and Computer Science}\\
                Department of Telecommunication Systems,
                Einsteinufer 27, 10587 Berlin, Germany\\
                \{\PAPERemailThree, \PAPERemailSeven, \PAPERemailEight \}@\PAPERemailTU
            }
            \vspace{0.05in}
            \IEEEauthorblockA{\IEEEauthorrefmark{3}
                \textit{NVIDIA}\\
                Fasanenstrasse 81, 10623 Berlin, Germany\\
                \{\PAPERemailFour, \PAPERemailFive, \PAPERemailNine\}@\PAPERemailNV
            }
            \vspace{-0.10in}
        }

    }
}
{}
}

\@ifclassloaded{article}  
{%
    \AtBeginDocument{%
        \title{\PAPERtitle}
        
        \author{%
                    \PAPERauthorOne $^1$
            \And    \PAPERauthorFour $^3$
            \And    \PAPERauthorTwo $^1$
            \And    \PAPERauthorThree $^1$
            \And    \PAPERauthorFive $^3$
            \And    \PAPERauthorSix $^1$
            \And    \PAPERauthorSeven $^1$
            \And    \PAPERauthorEight $^{1,2}$
            \And    \PAPERauthorNine $^3$        
        }
        \date{%
            $^1$\textit{Fraunhofer Institute for Telecommunications}, \textit{Heinrich Hertz Institute (HHI)}\\
            Department of Wireless Communications and Networks, Einsteinufer 37, 10587 Berlin, Germany\\[1ex]%
            $^2$\textit{Technical University Of Berlin}, \textit{Faculty IV - Electrical Engineering and Computer Science}\\
            Department of Telecommunication Systems, Einsteinufer 27, 10587 Berlin, Germany\\[1ex]%
            $^3$\textit{NVIDIA}, Fasanenstrasse 81, 10623 Berlin, Germany\\[2ex]%
            \today
        }
    }
}
{}

\makeatother


%% file: settings/hyperref.tex
\usepackage[]{hyperref}

\hypersetup{
  pdftitle    = {\PAPERtitle},
  pdfauthor   = {\PAPERauthorOne, \PAPERauthorFour, \PAPERauthorTwo, \PAPERauthorThree, \PAPERauthorFive, \PAPERauthorSix, \PAPERauthorSeven, \PAPERauthorEight \ and \PAPERauthorNine},
  pdfkeywords = {\PAPERkeywords},
  pdfsubject  = {IEEE Access journal paper},
  pdflang={english},
  hidelinks,                    
  pdfdisplaydoctitle=true,		
}


%% file: settings/cleveref.tex
\usepackage[capitalise,noabbrev]{cleveref}

\makeatletter
\newcommand*{\org@overidelabel}{}
\let\org@overridelabel\@verridelabel
\@ifpackagelater{acronym}{2015/03/21}{
  \renewcommand*{\@verridelabel}[1]{%
    \@bsphack
    \protected@write\@auxout{}{\string\AC@undonewlabel{#1@cref}}%
    \org@overridelabel{#1}%
    \@esphack
  }%
}{
  \renewcommand*{\@verridelabel}[1]{%
    \@bsphack
    \protected@write\@auxout{}{\string\undonewlabel{#1@cref}}%
    \org@overridelabel{#1}%
    \@esphack
  }%
}
\makeatother


%% file: settings/graphic.tex
\usepackage{graphicx}

\graphicspath{{./pictures/}{./figures/}}

\usepackage{support-caption}
\usepackage[labelformat=simple]{subcaption}

\DeclareCaptionLabelSeparator{periodspace}{.\quad}
\makeatletter
\@ifclassloaded{ieeeaccess}{%
\captionsetup{labelfont={color=accessblue}}
}
\makeatother

\makeatletter
\if@twocolumn
\newcommand{\whencolumns}[2]{
#2
}
\else
\newcommand{\whencolumns}[2]{
#1
}
\fi
\makeatother


%% file: settings/color.tex
\usepackage{xcolor}

\definecolor{HighlightColor}{rgb}{0.462745098, 0.725490196, 0.000000000}
\definecolor{nvfootlinecolor}{rgb}{0.55, 0.55, 0.55}
\definecolor{nvtextcolor}{rgb}{0.3686, 0.3686, 0.3686}
\definecolor{NVEmerald}{rgb}{0.0, 0.521568627, 0.392156863}
\definecolor{NVAmethyst}{rgb}{0.364705882, 0.08627451, 0.509803922}
\definecolor{NVCPUBlue}{rgb}{0.0, 0.443137255, 0.77254902}
\definecolor{NVGarnet}{rgb}{0.537254902, 0.047058824, 0.345098039}
\definecolor{NVFluorite}{rgb}{0.980392157, 0.760784314, 0.0}
\definecolor{NVLightGray}{rgb}{0.803921569, 0.803921569, 0.803921569}
\definecolor{NVMediumGray}{rgb}{0.549019608, 0.549019608, 0.549019608}
\definecolor{NVDarkGray}{rgb}{0.368627451, 0.368627451, 0.368627451}
\definecolor{NVRed}{rgb}{0.945, 0.349, 0.373} 

\definecolor{AmazonColor}{rgb}{1.000000000, 0.662745098, 0.000000000}

\definecolor{MATLAB_color1}{rgb}{0.0000, 0.4470, 0.7410}
\definecolor{MATLAB_color2}{rgb}{0.8500, 0.3250, 0.0980}
\definecolor{MATLAB_color3}{rgb}{0.9290, 0.6940, 0.1250}
\definecolor{MATLAB_color4}{rgb}{0.4940, 0.1840, 0.5560}
\definecolor{MATLAB_color5}{rgb}{0.4660, 0.6740, 0.1880}
\definecolor{MATLAB_color6}{rgb}{0.3010, 0.7450, 0.9330}
\definecolor{MATLAB_color7}{rgb}{0.6350, 0.0780, 0.1840}
\definecolor{MATLAB_color1_old}{rgb}{0.00, 0.00, 1.00}
\definecolor{MATLAB_color2_old}{rgb}{0.00, 0.50, 0.00}
\definecolor{MATLAB_color3_old}{rgb}{1.00, 0.00, 0.00}
\definecolor{MATLAB_color4_old}{rgb}{0.00, 0.75, 0.75}
\definecolor{MATLAB_color5_old}{rgb}{0.75, 0.00, 0.75}
\definecolor{MATLAB_color6_old}{rgb}{0.75, 0.75, 0.00}
\definecolor{MATLAB_color7_old}{rgb}{0.25, 0.25, 0.25}

\definecolor{MATLAB_blue}{rgb}{0,0,1}
\definecolor{MATLAB_black}{rgb}{0,0,0}
\definecolor{MATLAB_red}{rgb}{1,0,0}
\definecolor{MATLAB_green}{rgb}{0,1,0}
\definecolor{MATLAB_yellow}{rgb}{1,1,0}
\definecolor{MATLAB_cyan}{rgb}{0,1,1}
\definecolor{MATLAB_magenta}{rgb}{1,0,1}
\definecolor{MATLAB_white}{rgb}{1,1,1}

\definecolor{accessblue}{HTML}{00629B} 

\definecolor{greycolor}{HTML}{333333} 


%% file: settings/table.tex
\usepackage{tabularx}
\usepackage{colortbl}

\newcolumntype{L}{>{\raggedright\arraybackslash}X}
\newcolumntype{C}{>{\centering\arraybackslash}X}
\newcolumntype{R}{>{\raggedleft\arraybackslash}X}

\PassOptionsToPackage{table}{xcolor}

\usepackage{multirow}
\usepackage{hhline}



%% file: settings/tikz.tex
\usepackage{tikz}
\usepackage{pgfpages}
\usepackage{relsize}

\usetikzlibrary{
  fadings,
  graphs,
  shapes,
  arrows,
  fit,
  calc,
  positioning,
  external,
  arrows.meta,
  decorations.pathreplacing
}

\tikzset{
  myrotate/.style={rotate=#1,nodes={rotate=#1}}
}

\makeatletter
\newcommand{\gettikzxy}[3]{%
  \tikz@scan@one@point\pgfutil@firstofone#1\relax
  \edef#2{\the\pgf@x}%
  \edef#3{\the\pgf@y}%
}
\makeatother

\usepackage{pgfplots}
\DeclareUnicodeCharacter{2212}{−}
\usepgfplotslibrary{groupplots,dateplot}
\pgfplotsset{compat=newest}


%% file: settings/siunix.tex
\usepackage{siunitx}

\DeclareSIUnit{\sample}{Sample}

\DeclareSIQualifier\isotropic{i}



%% file: sections/abstract.tex


\begin{abstract}
In this feasibility study, we have implemented a recently proposed partially linear multiuser detection algorithm in \acp{RKHS} on a \acs{GPU}-accelerated platform.
Partially linear multiuser detection, which combines the robustness of linear detection with the power of nonlinear methods, has been proposed for a massive connectivity scenario with the \ac{NOMA}.
This is a promising approach, but detecting payloads within a received \ac{OFDM} radio frame requires the execution of a large number of inner product operations, which are the main computational burden of the algorithm. 
Although inner-product operations consist of simple kernel evaluations, their vast number poses a challenge in \ac{ULL} applications, because the time needed for computing the inner products might exceed the sub-millisecond latency requirement.
To address this problem, this study demonstrates the acceleration of the inner-product operations through massive parallelization.
The result is a \acs{GPU}-accelerated real-time \ac{OFDM} receiver that enables sub-millisecond latency detection to meet the requirements of \ac{5G} and beyond \ac{URLLC} systems.
Moreover, the parallelization and acceleration techniques explored and demonstrated in this study can be extended to many other signal processing algorithms in Hilbert spaces, such as those based on \ac{POCS} and \ac{APSM} algorithms. 
Experimental results and comparisons with the state-of-art confirm the effectiveness of our techniques.

\end{abstract}

\comment{
\begin{abstract}
Adaptive partial linear beamforming meets the need of \acs{5G} and future \acs{6G} applications for high flexibility and adaptability.
Choosing an appropriate tradeoff between conflicting goals opens the recently proposed \ac{MU} detection method.
Due to their high spatial resolution, nonlinear beamforming filters can significantly outperform linear approaches in stationary scenarios with massive connectivity.
However, a dramatic decrease in performance can be expected in high mobility scenarios because they are very susceptible to changes in the wireless channel.
The robustness of linear filters is required, considering these changes.
One way to respond appropriately is to use online machine learning algorithms.
The theory of algorithms based on the \ac{APSM} is rich, and they promise accurate tracking capabilities in dynamic wireless environments.
However, one of the main challenges comes from the real-time implementation of these algorithms, which involve projections on time-varying closed convex sets.
While the projection operations are relatively simple, their vast number poses a challenge in \ac{ULL} applications where latency constraints must be satisfied in every radio frame.
Taking \ac{NOMA} systems as an example, this paper explores the acceleration of \ac{APSM}-based algorithms through massive parallelization.
The result is a \acs{GPU}-accelerated real-time implementation of an \ac{OFDM}-based transceiver that enables detection latency of less than one millisecond and therefore complies with the requirements of 5G and beyond.
To meet the stringent physical layer latency requirements, careful co-design of hardware and software is essential, especially in virtualized wireless systems with hardware accelerators.
\end{abstract}
}

\comment{
\begin{abstract}
Adaptive partial linear beamforming, as recently proposed for \ac{MU} detection, opens up the possibility of trading off the high spatial resolution of nonlinear beamforming filters for the robustness of linear filters to changes in dynamic wireless environments.
The ability to choose an appropriate tradeoff between conflicting goals such as robustness and spatial resolution meets the need of \acs{5G} and future \acs{6G} applications for high flexibility and adaptability.
For example, nonlinear beamforming can significantly outperform linear approaches in stationary massive connectivity scenarios, while in high mobility scenarios the performance of nonlinear filters can degrade dramatically because, unlike linear filters, they are very sensitive to temporal changes in the wireless channel.
Dealing with these changes requires the use of online machine learning algorithms with good tracking capabilities, such as those based on the \ac{APSM}.
The theory of \ac{APSM}-based algorithms is rich and they promise very good tracking capabilities in dynamic wireless environments.
However, one of the key challenges comes from the real-time implementation of these algorithms, which involve projections on time-varying closed convex sets.
While the projection operations are relatively simple, their vast number poses a serious challenge in \ac{ULL} applications where  given latency constraints must be satisfied in every frame.
Taking \ac{NOMA} systems as an example, this paper explores the acceleration of \ac{APSM}-based algorithms through massive parallelization.
We show the processing delay optimization of the detection phase functions by using parallel running threads on a \ac{GPU} because these functions are also reused in the training phase.
The result is a \ac{GPU}-accelerated real-time implementation of an \ac{OFDM}-based transceiver that enables detection latency of less than one millisecond and therefore complies with the requirements of \acs{5G} and beyond.
This paper clearly demonstrates - and this is a key message of this study - that careful co-design of hardware and software is essential to meet the stringent physical layer latency requirements in virtualized wireless systems utilizing hardware accelerators.
\end{abstract}
}

\comment{
\begin{abstract}
Adaptive partially linear filters as recently proposed for \ac{MU} detection offer robustness against small changes in dynamic wireless environments. 
They can take advantage of situations where linear filters achieve near-optimal performance and otherwise outperform linear filters due to their \acl{NL} components.
The online machine learning algorithm, which is a special case of the \ac{APSM}, mainly consists of projections onto time-varying closed-convex sets.
While the operations are simple, their mere number is a challenge in \ac{ULL} applications, where real-time operation and guaranteed latency are a requirement.
We consider the example \ac{NOMA} systems, to explore the acceleration of the projection-based method and develop a massively parallel algorithm resulting in a \ac{GPU}-accelerated real-time implementation that is suitable for practical communication systems.
We demonstrate that our proof-of-concept of an \ac{OFDM}-based transceiver setup is capable of a detection latency of less than one millisecond, which is in line with 5G and beyond requirements.
\end{abstract}
}


%% file: sections/I_introduction.tex

\section{Introduction}
\label{sec:Intro}
Recently, a large body of research has been devoted to \ac{NOMA} \cite{Shin2017,Wang2016,Ding2017,Tabassum2016} because the requirements of massive connectivity beyond \ac{5G} mobile networks necessitate the efficient use of time-frequency resources.
In contrast to traditional \ac{OMA}, such as \ac{OFDMA}, \ac{NOMA} allocates the same time-frequency resource to multiple users in the same cell, which may result in strong multiuser interference.
To deal with the strong interference, partially linear multiuser detection in \acp{RKHS} has been proposed (\eg, see \cite{HIGUCHI2015,Xin2016,Islam2017,9148869,8422449,Awan_thesis}).
The reason for considering partially linear receivers is that, while nonlinear detectors can outperform linear detectors significantly in scenarios with strong multiuser interference, they may be highly sensitive even to small changes in a wireless environment (\eg, those caused by intermittent interference and multipath scattering in \ac{MMTC} scenarios).
In fact, theoretical studies \cite{Bjornson2017} have shown that linear detectors can achieve a comparable spectral efficiency to nonlinear methods in massive \ac{MIMO} systems.
Therefore, in massive \ac{MIMO} \ac{NOMA} systems, purely nonlinear detectors may be inefficient.
Based on these facts, studies in \cite{9148869,8422449,Awan_thesis} have designed a hybrid multiuser detector in \acp{RKHS} that combines the strengths of nonlinear and linear filters.
The proposed supervised projection-based learning algorithm, which is a special case of the \ac{APSM} \cite{Yamada}, has some very desirable features.
For example, the algorithm learns to detect modulation symbols of a user directly without any intermediate parameter estimation which is often prone to errors.
Furthermore, if the channel information is available, the algorithm can be initialized by a conventional linear filter, and the performance can be further improved in scenarios for which linear filters are insufficient when dealing with strong interference. 

Basically, the \ac{APSM}-based algorithm tracks the intersection of time-varying closed-convex sets to which the desired solution (often a vector or a function) belongs.
In this regard, it resembles a classic projection-based set-membership estimation\footnote{%
It is important to mention that the classical set-membership estimation considers a finite number of sets whereas in our \acs{APSM}-based algorithm an infinite number of sets is considered.
}%
(\eg, see \cite{Combettes93,Stark1998}). In general, such algorithms consist of a sequence of projections on the constructed closed convex subsets of an appropriately defined Hilbert space.
The closed convex subsets are often chosen such that the projections (which require computing inner products) admit a closed-form solution; for example, projections onto hyperplanes, closed balls, closed halfspaces, \etc.
It is known that the computational effort of projection-based algorithms is dominated by the cost of computing projections.
In \ac{ULL} applications, however, straightforward implementations of a large number of seemingly simple inner products in an \ac{OFDM} radio frame may consume too much time to satisfy the temporal and latency constraints. 
Luckily, many algorithms that involve projections and inner products in Hilbert spaces lend themselves very well to massive parallelism.
In order to realize our goal of real-time \acs{GPU}-accelerated multiuser detection, we exploit the intrinsic parallelism of \ac{APSM} and carefully schedule operations across the memory hierarchy of the \ac{GPU}. 

\comment{
In the future, various \ac{RAN} functions will be implemented in software and, where possible, run efficiently and with low latency on general-purpose hardware.
The \ac{RAN} functions will have the ability to simultaneously access these computing resources, which are available at different locations.
Considering the available computing power and energy and cost constraints, the virtualization of RAN functions is challenging, especially in \acp{DU}, which does the most computational work in baseband processing.
}

Before we move on to our contributions, we mention the fact that in the future many \ac{RAN} functions will be virtualized on general-purpose \ac{COTS} hardware. Recent industry trials have shown that virtualization of \ac{RAN} functions in the \acp{DU}, consisting of computationally intensive baseband operations, may not be cost-effective (compared to traditional \ac{ASIC} hardware) and viable without using modern acceleration platforms. Therefore, baseband algorithms that are suited to such platforms are the subject of current interest in the wireless industry.

\subsection{Contributions}
Our main contribution is a massively parallel GPU-accelerated \ac{NOMA} system that can fulfill the real-time constraints of \ac{ULL} in an \ac{OFDM}-based system.
Below, we summarize our contributions.
\begin{itemize}
\item Our focus in this study is to accelerate the detection of an \ac{OFDM} radio frame which may involve computations of a large number of inner products.
The acceleration is achieved by state-of-art parallelization and memory management techniques.  
\item The performance of the developed \acs{GPU}-accelerated system is then demonstrated in a setup with 6 transmit and 16 receive antennas.
Compared to the existing \ac{PoC} MATLAB implementation \cite{9148869} we demonstrate that using massively parallel processor structures on a \ac{GPU} reduces the processing time to milliseconds.
\item \acs{GPU}-accelerated signal processing is an important part of \ac{5G} and beyond communication systems \cite{nvidia_web_gpu}.
Such platforms are particularly suited to signal processing algorithms consisting of a large number of projections requiring inner products that can be executed in parallel.
This means that the techniques that we have developed in this work can also be used in other inner product (and projection) based algorithms, such as projection onto convex sets (POCS) signal processing in Hilbert spaces. 
\end{itemize}

\subsection{Structure}
The paper is organized as follows.
In \cref{sec:Background}, we review the algorithm for multiuser detection proposed in \cite{8422449}.
In particular, some practical aspects of the algorithm are explained along with its potential for acceleration by parallel signal processing.
\Cref{sec:Implementation} discusses the implementation of the algorithm and the techniques to optimize parallel processing and memory access on the targeted \ac{GPU} platform.
\Cref{sec:SystemSetup} describes the hardware equipment and the experimental setup deployed in the proof-of-concept.
Finally, the results are presented in \cref{sec:Res}.
 

%% file: sections/II_background.tex

\section{Machine learning based Multiuser Detection}
\label{sec:Background}
In this section, we describe the \acs{APSM}-based algorithm for multiuser detection that has been studied in \cite{9148869,8422449,Awan_thesis}. Here, we focus on the implementation and computational aspects of the algorithm, and we avoid repeating unnecessary details. 
\subsection{Preliminaries}
\label{sec:preliminaries}
In the following, the sets of natural numbers, non-negative integers, real numbers, and complex numbers are denoted by $\mathbb{N}$, $\mathbb{N}_0$, $\mathbb{R}$, and $\mathbb{C}$, respectively.
We define the range $\overline{N_{1},N_{2}}:=\left\{N_1,N_{1}+1,\ldots,N_2\right\}$, where $N_1, N_2 \in \mathbb{N}_0$ and $N_1\leq N_2$. 

This study deals with adaptive filtering in special (real) Hilbert spaces known as \acp{RKHS}.
Briefly, given an arbitrary (linear) subspace $\mathcal{U}\subseteq \mathbb{R}^{l}$, an \ac{RKHS} $(\mathcal{H},\left\langle \cdot,\cdot \right\rangle_{\mathcal{H}})$ is a Hilbert space with the inner product $\left\langle \cdot,\cdot \right\rangle_{\mathcal{H}}$, uniquely associated with a positive definite function known as the reproducing kernel $\kappa:\mathcal{U}\times\mathcal{U}\rightarrow \mathbb{R}$ of $\mathcal{H}$.
In this study, $\kappa$ is either the linear kernel denoted by  $\kappa_{\mathrm{L}}$ and defined as $(\forall\mathbf{u}, \mathbf{v} \in\mathcal{U})$ $\kappa_{\mathrm{L}}(\mathbf{u},\mathbf{v}):=\mathbf{u}^{T}\mathbf{v}$ or the Gaussian kernel denoted by $\kappa_{\mathrm{G}}$ and defined as $(\forall\mathbf{u}, \mathbf{v} \in\mathcal{U})$ $\kappa_{\mathrm{G}}(\mathbf{u},\mathbf{v}):=\exp\left(-\frac{\|\mathbf{u}-\mathbf{v}\|^{2}_{\mathbb{R}^{l}}}{2\sigma^{2}}\right)$, where $\|\cdot\|_{\mathbb{R}^{l}}$ is the Euclidean norm in $\mathbb{R}^{l}$ and the variance $\sigma^{2}>0$ is a design parameter for the kernel width.
The \acp{RKHS} associated with $\kappa_{\mathrm{L}}$ and $\kappa_{\mathrm{G}}$ are denoted by $(\mathcal{H}_{\mathrm{L}}, \left\langle \cdot,\cdot \right\rangle_{\mathcal{H}_{\mathrm{L}}})$ and $(\mathcal{H}_{\mathrm{G}}, \left\langle \cdot,\cdot \right\rangle_{\mathcal{H}_{\mathrm{G}}})$, respectively. 

Now consider either of the above-mentioned \acp{RKHS} and the well-known property that $(\forall\mathbf{u}\in\mathcal{U})$ $\kappa(\mathbf{u},\cdot)\in\mathcal{H}$.
In this study, we deal with functions of the type $f:=\sum_{n=1}^{N}a_n\kappa(\mathbf{u}_n,\cdot) \in \mathcal{H}$, where $a_n\in\mathbb{R}$ and $\mathbf{u}_n \in \mathcal{U}$.
\comment{Note that all possible linear combinations of $\kappa(\mathbf{u},\cdot)$ are contained in $\mathcal{H}$ because $\mathcal{H}$ is a Hilbert space.}
For two functions $f:=\sum_{n=1}^{N}a_n\kappa(\mathbf{u}_n,\cdot) \in \mathcal{H}$ and $g:=\sum_{m=1}^{M}b_m\kappa(\mathbf{v}_m,\cdot)  \in \mathcal{H}$, where $a_m,b_m\in\mathbb{R}$ and $\mathbf{u}_n,\mathbf{v}_m \in \mathcal{U}$, we define the inner product
\begin{equation}
    \left\langle f,g\right\rangle_{\mathcal{H}}:= \sum_{n=1}^{N}\sum_{m=1}^{M}a_n b_m\kappa(\mathbf{u}_n,\mathbf{v}_m),
    \label{eqn:inner_product_definition}
\end{equation}
inducing the norm
\begin{equation}
    \left\|f\right\|^{2}_{\mathcal{H}}=\left\langle f,f\right\rangle_{\mathcal{H}}.      
\end{equation}

\subsection{Partially Linear Filter Design}
The studies in \cite{9148869,8422449,Awan_thesis} combine the robustness of linear beamforming filters with a higher spatial resolution of nonlinear beamforming filters by designing a partially linear filter in \acp{RKHS}.
In more detail, the \ac{RKHS} $\mathcal{H}_{\mathrm{G}}$ associated with the Gaussian kernel $\kappa_{\mathrm{G}}$ is combined (\ie, summed) with the \ac{RKHS} $\mathcal{H}_{\mathrm{L}}$ associated with the linear kernel $\kappa_{\mathrm{L}}$ to obtain a sum \ac{RKHS} of partially linear filters.
To be more precise, a partially linear filter $f$ is defined as an element of the real \ac{RKHS} $\mathcal{H}:=\mathcal{H}_{\mathrm{L}} + \mathcal{H}_{\mathrm{G}}:=\left\{w_{\mathrm{L}}f_{\mathrm{L}}+w_{\mathrm{G}}f_{\mathrm{G}} : f_{\mathrm{L}}\in\mathcal{H}_{\mathrm{L}}, f_{\mathrm{G}}\in\mathcal{H}_{\mathrm{G}}\right\}$, where $w_{\mathrm{L}},w_{\mathrm{G}} \geq 0$ are fixed weights for the linear and the Gaussian part, respectively.
Fact~\ref{fact:mf} shows how the kernel and inner products are computed in $\mathcal{H}$:

\begin{fact}[\textit{Reproducing kernel of the weighted sum space \cite{Yukawa2015}}]
\textit{Assume that the input space $\mathcal{U}\subseteq\mathbb{R}^{l}$ has a nonempty interior.
Then, given any $w_{\mathrm{L}},w_{\mathrm{G}}>0$ and $\mathbf{u},\mathbf{v}\in\mathcal{U}$,  $\kappa(\mathbf{u},\mathbf{v}):=w_{\mathrm{L}}\kappa_{\mathrm{L}}(\mathbf{u},\mathbf{v})+w_{\mathrm{G}}\kappa_{\mathrm{G}}(\mathbf{u},\mathbf{v})$ is the reproducing kernel of the sum space $\mathcal{H}$ equipped with the inner product}
\begin{equation}
    \left\langle f,g\right\rangle_{\mathcal{H}}:=w_{\mathrm{L}}^{-1}\left\langle f_{\mathrm{L}},g_{\mathrm{L}}\right\rangle_{\mathcal{H}_{\mathrm{L}}}+
                                                 w_{\mathrm{G}}^{-1}\left\langle f_{\mathrm{G}},g_{\mathrm{G}}\right\rangle_{\mathcal{H}_{\mathrm{G}}}.
    \label{eqn:sum_inner_product}
\end{equation}
\label{fact:mf}
\end{fact}
The inner products in the two component \acp{RKHS} in \eqref{eqn:sum_inner_product} are simple to compute because, according to \eqref{eqn:inner_product_definition}, they consist of (mainly) kernel evaluations. Note that the closed form in \eqref{eqn:sum_inner_product} is in general not valid for arbitrary choices of the two kernels. 

\subsection{Multiuser Detection}
\label{sec:multiuser_detection}
Consider a multiuser uplink with $K$ users and $M$ receive antennas. We assume a non-dispersive channel so that the received signal (sampled at a fixed symbol rate) at the time $t \in \mathbb{N}_0$ is given by 
\begin{align}
	\mathbf{r}:\mathbb{N}_0\rightarrow \mathbb{C}^{M}:t\mapsto&\left[{r}_{1}(t),{r}_{2}(t),\ldots,{r}_{M}(t)\right]^{\intercal}\\
		           =&\sum_{k=1}^{K}\sqrt{p_{k}}b_{k}(t)\mathbf{h}_{k}+ \mathbf{n}(t),
	\label{eqn:uplink_signal}
\end{align} 
where $p_{k}\in\mathbb{R}$ is the transmit power of user $k \in \overline{1,K}$, and $b_{k}(t)\in\mathbb{C}$ is the modulation symbol. The vectors $\mathbf{h}_{k}\in\mathbb{C}^M$ and $\mathbf{n}(t)\in\mathbb{C}^M$ stand for the channel signature of user $k$ and additive noise, respectively.
Note that we do not assume any distribution and structure of the noise and the receive antenna array, respectively.
The objective of multiuser detection considered in this study is to design a  filter $g^{k}:\mathbb{C}^{M}\rightarrow\mathbb{C}$ for a selected user $k$, such that $(\forall t\in\mathbb{N}_0)$ $\left|g^{k}(\mathbf{r}(t))-{b}_{k}(t)\right|\leq\epsilon$, where $\epsilon>0$ is a small predefined noise tolerance.
In other words, the goal is to detect the desired modulation symbols directly without any intermediate parameter estimation. 

\subsection{Adaptive Detection in Sum \acs{RKHS}}
\label{sec:apsm}
In this section, we describe the \acs{APSM}-based detection algorithm of \cite{9148869,8422449,Awan_thesis}. We convert the complex vector $\mathbf{r}(t)\in\mathbb{C}^{M}$ into two real vectors $\mathbf{r}_{1}(t):=\left[\Re(\mathbf{r}(t))^{\intercal}\,\Im(\mathbf{r}(t))^{\intercal}\right]^{\intercal}\in\mathbb{R}^{2M}$ and $\mathbf{r}_{2}(t):=\left[\Im(\mathbf{r}(t))^{\intercal}\, -\Re(\mathbf{r}(t))^{\intercal}\right]^{\intercal}\in\mathbb{R}^{2M}$ which enables processing in real Hilbert spaces as considered in \cite{Yamada2005,Slavakis2009}. Similarly, the training modulation symbols are converted to $\left[b_{1}(t)\,b_{2}(t)\right]^{\intercal}:=\left[\Re(b(t)) \,\Im(b(t))\right]^{\intercal} \in \mathbb{R}^2$. The proposed filter $f:\mathbb{R}^{2M}\rightarrow\mathbb{R}$ operates on $\mathbf{r}_{1}(t)$ and $\mathbf{r}_{2}(t)$ separately. The relation between $f$ and the complex-valued filter $g$ described in \cref{sec:multiuser_detection} is given by $(\forall t\in\mathbb{N}_0)$ $(\mathbb{C} \ni) g(\mathbf{r}(t))=f(\mathbf{r}_{1}(t))+if(\mathbf{r}_{2}(t))$, where $i$ is the solution to the equation $i^2=-1$.
To simplify indexing, we define a new time index: $(\forall t\in\mathbb{N}_0)$ $(\forall{l}\in\overline{1,2})$ $n:=2t+l-1$, $\mathbf{r}_{n}=\mathbf{r}_{2t+l-1}:=\mathbf{r}_{l}(t)$ and $b_{n}=b_{2t+l-1}:=b_{l}(t)$.
Henceforth, we denote the input space of received signals by $\mathcal{U}:=\left\{\mathbf{r}_n\in\mathbb{R}^{2M}:n\in\mathbb{N}_0\right\}$.
We now turn our attention to the design of an adaptive filter $f$ such that $(\forall n\in\mathbb{N}_0)$ $|f(\mathbf{r}_n)-b_n|\leq \epsilon$, where the precision is controlled by the design parameter $\epsilon>0$.
We assume that $f\in\mathcal{H}$ and a training sample $(\mathbf{r}_n,b_n)\in\mathcal{U} \times \mathbb{R}$ is available $\forall n\in\mathbb{N}_0$.
Then, a closed and convex set of functions in $\mathcal{H}$ consistent with the training sample at time $n$ is given by 
\begin{equation}
    C_n:=\left\{f\in\mathcal{H}:|\left\langle f,\kappa(\mathbf{r}_n,\cdot)\right\rangle_{\mathcal{H}}-b_n|\leq \epsilon\right\}.  
    \label{eqn:convex_set}
\end{equation}
In the online learning setting considered here, the training samples arrive as a sequence and each sample defines a set of the form in \eqref{eqn:convex_set}.
Ideally, the objective is to find a filter $f^{\ast}\in\mathcal{H}$ such that $f^{\ast}$ is a member of all these sets, \ie, $f^{\ast}\in \bigcap_{n\in\mathbb{N}_0}C_n$. 
However, since it is challenging to find a low-complexity algorithm to solve this problem, we allow a finite number of sets not to share a common intersection and consider the simplified problem:

\begin{equation}
    \text{find}\, f^{\ast} \in \bigcap_{n\geq n_o}C_n,
    \label{eqn:optimization problem}
\end{equation}
for some $n_o\in\mathbb{N}_0$, under the assumption that $\bigcap_{n\geq n_o}C_n \neq \emptyset$.
The advantage of the above formulation is that we can find an $f\in\mathcal{H}$ that is close to the intersection in \eqref{eqn:optimization problem} utilizing an \acs{APSM}-based \cite{Yamada2005,Theodoridis2011} algorithm which we describe below. 

As in \cite{Slavakis2009,Theodoridis2011}, given an index set $\mathcal{J}_n$ of size $W$, and starting from an arbitrary $f_0 \in \mathcal{H}$, we construct a sequence of filter estimates in $\mathcal{H}$ given as
\begin{equation}
    (\forall n\in\mathbb{N}_0)\, f_{n+1}=\sum_{j\in\mathcal{J}_n}q^{n}_j\mathbf{P}_{C_j}(f_n),
    \label{eqn:apsm}
\end{equation}
where $\mathbf{P}_{C_j}(f_n)=f_n+\beta^n_j\kappa(\mathbf{r}_{j},\cdot)=f_n+\beta^n_j(w_{\mathrm{L}}~\kappa_{\mathrm{L}}(\mathbf{r}_{j},\cdot)+w_{\mathrm{G}}~\kappa_{\mathrm{G}}(\mathbf{r}_{j},\cdot))$ is the projection of $f_{n}$ onto the set $C_j$, with $\beta^n_j$ given by 
\begin{align}
	\beta^{n}_{j}:= \begin{cases} 
  \frac{b_j-\left\langle f_n,\kappa(\mathbf{r}_j,\cdot)\right\rangle_{\mathcal{H}}\,-\epsilon}{\kappa(\mathbf{r}_j,\mathbf{r}_j)}, &\text{if }  \left\langle f_n,\kappa(\mathbf{r}_j,\cdot)\right\rangle_{\mathcal{H}}-b_j < -\epsilon,\\
  0, & \text{if } |\left\langle f_n,\kappa(\mathbf{r}_j,\cdot)\right\rangle_{\mathcal{H}}-b_j| \leq \epsilon,\\
  \frac{b_j-\left\langle f_n,\kappa(\mathbf{r}_j,\cdot)\right\rangle_{\mathcal{H}}\,+\epsilon}{\kappa(\mathbf{r}_j,\mathbf{r}_j)}, & \text{if }  \left\langle f_n,\kappa(\mathbf{r}_j,\cdot)\right\rangle_{\mathcal{H}}-b_j > \epsilon,
	\end{cases}
	\label{eqn:beta}
\end{align}
and where $(q^n_j)_{j\in\mathcal{J}_n}$ are non-negative weights satisfying $\sum_{j\in\mathcal{J}_n}q^{n}_j=1$.

The index set $\mathcal{J}_n$ defined as $\mathcal{J}_{n}:=\overline{n-W+1,n}$ if $n \geq W-1$, otherwise $\mathcal{J}_{n}:=\overline{0,n}$, allows for a subset of sets $C_1,C_2,\ldots,C_n$ to be processed concurrently to accelerate convergence, and the weights ${q}^{n}_{j}$ can be used to adaptively prioritize the sets.
The computational advantage of this algorithm is that the projection $\mathbf{P}_{C_j}(f_n)$ only requires simple inner products, and the overall algorithm is amenable to parallelization resulting in significant acceleration as discussed in \cref{sec:acceleration_via_parallel_processing}. 

Before we move onto the next section, it can be verified that the filter estimate generated by \eqref{eqn:apsm} can be decomposed as $f_{n}:=\sum_{i=1}^{n-1}\gamma^{(n)}_{i}\kappa(\mathbf{r}_i,\cdot)=\sum_{i=1}^{n-1}\gamma^{(n)}_{i}w_{\mathrm{L}}\kappa_{\mathrm{L}}(\mathbf{r}_i,\cdot)+\sum_{i=1}^{n-1}\gamma^{(n)}_{i}w_{\mathrm{G}}\kappa_{\mathrm{G}}(\mathbf{r}_i,\cdot)=:f_{\mathrm{L},n}+f_{{\mathrm{G}},n}$, where $f_{\mathrm{L},n} \in \mathcal{H}_{\mathrm{L}}$ and $f_{\mathrm{G},n} \in \mathcal{H}_{\mathrm{G}}$ \cite{8422449}.
Since $\mathcal{H}_{\mathrm{L}}$ is nothing but the Euclidean space $\mathbb{R}^{2M}$, it is spanned by the Euclidean basis (which we refer to as the \textit{linear dictionary} in the following) $\mathcal{D}_{\mathrm{L}}:=\left\{w_{\mathrm{L}}\kappa_{\mathrm{L}}(\mathbf{e}_1,\cdot), w_{\mathrm{L}}\kappa_{\mathrm{L}}(\mathbf{e}_2,\cdot),\ldots,w_{\mathrm{L}}\kappa_{\mathrm{L}}(\mathbf{e}_{2M},\cdot)\right\}$, where $\mathbf{e}_m\in\mathbb{R}^{2M}$ is the canonical basis vector having a one at the $m$-th index and zeros elsewhere.
So, every $\kappa_{\mathrm{L}}(\mathbf{r}_n,\cdot)$ can be expressed as $\sum_{m=1}^{2M}[\mathbf{r}_n]_{m}\kappa_{\mathrm{L}}(\mathbf{e}_m,\cdot)$, with $[\mathbf{r}_n]_{m}$ the $m$-th entry of $\mathbf{r}_n$.
As a result, the linear component $(\forall n\in\Natural)~f_{\mathrm{L},n}=w_{\mathrm{L}}~\sum_{i=1}^{n-1}\gamma^{(n)}_{i}\kappa_{\mathrm{L}}(\mathbf{r}_i,\cdot) =w_{\mathrm{L}}~\sum_{m=1}^{2M}\gamma^{(\mathrm{L},n)}_{m}\kappa_{\mathrm{L}}(\mathbf{e}_m,\cdot)$ consists of $2M$ components with their coefficients $\gamma^{(\mathrm{L},n)}_{m}$ updated by projections $\mathbf{P}_{C_j}(f_n)$ in \eqref{eqn:apsm}.
In contrast to the linear component, the \textit{Gaussian dictionary} given by $\mathcal{D}_{\mathrm{G},n}:=\left\{w_{\mathrm{G}}\kappa_{\mathrm{G}}(\mathbf{r}_1,\cdot)\right.,\allowbreak w_{\mathrm{G}}\kappa_{\mathrm{G}}(\mathbf{r}_2,\cdot),\ldots,\left.w_{\mathrm{G}}\kappa_{\mathrm{G}}(\mathbf{r}_{n-1},\cdot)\right\}$ grows with time $n$ as the iterations progress.

\comment{
\begin{remark}[Combination with conventional techniques]\label{rem:combination_with_conventional_techniques}
Note that $f_0$ in \eqref{eqn:apsm} is arbitrary. Conventional receivers often work with explicit parameter knowledge (such as user channels) and linear filters. So, in principle, $f_0$ can be a conventional linear filter, i.e., $f_0 \in \mathcal{H}_{\mathrm{L}}$. It can be verified that, if $f_0 \notin \bigcap_{n \geq n_0} C_n$, subsequent iterations will seek to improve the performance. Otherwise, $f_0$ will be returned as the final estimate. This is particularly useful in cases where the initial purely linear filter cannot deal with excessive multiuser interference.
\end{remark}}
\begin{remark}[ML: Computation and Communication]\label{rem:ml_comm}
It is common in literature to assume that the channels between the users and the base stations remain constant only for a certain time period known as the \textit{channel coherence time} \cite{Du2017}. Therefore, the training time should be much smaller than the coherence time such that a large portion of the coherence time can be used for detection/communication. From a processing and computation point-of-view, the training time is the total time it takes to collect/sample the training set and learn a good detection filter. This means that learning algorithms that have low complexity and can be accelerated are highly desired. Furthermore, in high-data rate communication fast detection is also highly desirable, so that as much data as possible can be detected during the coherence time. 
\end{remark}

\subsection{Acceleration via Parallel Processing}
\label{sec:acceleration_via_parallel_processing}
As mentioned above, the concurrent projections in \eqref{eqn:apsm} accelerate the convergence.
Unfortunately, large values of $W$ result in significant computational and memory burden and this may result in intolerable latency in real-time applications. However, note that the $W$ projections in \eqref{eqn:apsm} are independent, which means that they can be computed in parallel on platforms equipped with \acp{GPU} optimized for such applications. In addition, the computation of $\beta^{n}_{j}$ in \eqref{eqn:beta} involves the inner product $\left\langle f_n,\kappa(\mathbf{r}_j,\cdot)\right\rangle_{\mathcal{H}}$. Since $f_{n}:=\sum_{i=1}^{n-1}\gamma^{(n)}_{i}\kappa(\mathbf{r}_i,\cdot)$, this inner product is a sum of $n$ independent inner products (equivalently kernel evaluations due to \eqref{eqn:inner_product_definition}), so these can also be computed in parallel. Furthermore, each linear kernel evaluation requires the dot-product $(\forall \mathbf{u},\mathbf{v} \in \mathcal{U})$ $\mathbf{u}^\intercal\mathbf{v}=\sum^{2M}_{i=1} [\mathbf{u}]_i[\mathbf{v}]_i$, while the Gaussian kernel evaluation requires the calculation of the Euclidean-norm $\sqrt{(\mathbf{u}-\mathbf{v})^\intercal(\mathbf{u}-\mathbf{v})}$ followed by further operations. This means that both kernel evaluations are independent component-by-component computations for any two input vectors followed by further operations on the sum or accumulation of these computations. The complexity of the above operations is linear in the dimension of the input vectors (the number of antennas $M$ in this study), however the computation time can be reduced drastically if each component-by-component computation is executed in parallel. Finally, each computation requires access to data stored in various memory locations (discussed in \cref{sec:Implementation}) on the computing platform which means that, in addition to the computational aspects, special attention should be paid to the memory allocation and access. Before we move on to the real-time implementation of the algorithm in \eqref{eqn:apsm}, we remark how this \ac{APSM}-based algorithm can be implemented in practical communication systems.
\begin{remark}[Focus on Detection]\label{sec:training_tracking_detection}
\ac{APSM} is an online algorithm that keeps track of the set of minima of an infinite number of time-changing objective functions. As mentioned in Remark \ref{rem:ml_comm}, in a traditional communication system, generally an initial ``training phase'' is carried out during which the desired user sends a certain number of training pilots to the receiver, which allows the receiver to approximate a good initial receive filter using \ac{APSM}. The training phase is then stopped, and it is followed by the communication or ``detection phase'' during which the response of the trained filter is used as an estimate of the desired user's modulation symbols. In principle, retraining is only required if the communication channels change abruptly resulting in large errors. In situations, such as mobile scenarios, where communication channels change gradually, we can, in principle, track these changes using APSM through the data symbols. Therefore, after the initial training phase, the latency is mainly dependent on the time incurred during detection of radio frames, which involves computation of a large number of inner products according to \eqref{eqn:inner_product_definition} and \eqref{eqn:sum_inner_product}. Due to this reason, the acceleration of detection is the main focus of our work in the following.  \end{remark}


%% file: sections/III_implementation.tex

\section{Real-time implementation}\label{sec:Implementation}
In this section, we review the target platform, followed by a detailed description of the parallelization and optimization techniques used to achieve real-time performance. 

\subsection{Platform overview}
\label{sec:cuda_overview}
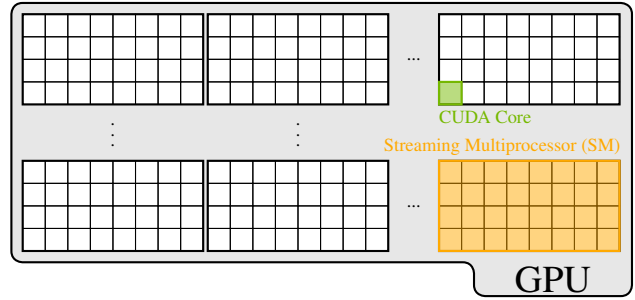
\begin{figure}[t] 
    %
    %
    \centering
    \scriptsize\centering\input{figures/tikz/cuda-device-mapping.tikz}
    \caption[mapping]{\acs{CUDA}-device mapping. A \acs{GPU} consists of
    thousands of \acs{CUDA} cores grouped as \acp{SM} executing code in \ac{SIMT} style. The number of \acp{SM} is determined by the size of the device. See \cref{sec:cuda_overview}.}
    \label{fig:cuda_device_mapping}
    %
    %
\end{figure}

\acp{GPU} are massively parallel processing devices that share a common architecture for general-purpose programming using the \ac{CUDA} programming model. \ac{CUDA} is a C++ extension that allows parallel code to be described as threads organized in a hierarchical structure, while also providing a framework for the communication between the host and the device. Each thread represents a logically independent sequence of instructions, similar to a thread in a classic programming model.
Threads are grouped into blocks (groups of threads) and grids (groups of blocks).
This hierarchical organization maps very well onto the device structure, with threads organized within the \acp{SM} as warps on the hardware cores. A warp is a common grouping of cores executing in parallel as \ac{SIMT}, and a \acs{SM} can have one or more warps active at any given time, depending on the number of cores available per \acs{SM} architecture. \acp{GPU} are composed of one or more \acp{SM} as depicted in \cref{fig:cuda_device_mapping}.

The above-mentioned hierarchy implies several types of memory, which differ in their sizes, bandwidths, and latencies. Global Memory is available to all threads and it is the largest (several Gigabytes off chip), but it also has the highest latency (from a hundred up to thousands of cycles). Shared Memory is available within \acp{SM}, and it is only available to threads belonging to the same block.
It has lower latency (tens of cycles) than Global Memory and is smaller in size (a few kilobytes on chip). Finally, registers are local to every thread, small in number (a few per thread in a typical scenario) with the lowest latency.

Note that threads can cooperate with each other by collaborating to perform tasks more efficiently or operating in batches, for example, to reduce the perceived latency of the overall memory access (known as latency hiding). Finally, the architecture has limited coherence of both execution and memory consistency at different levels of the hierarchy. A detailed description of the platform is found in the \ac{CUDA} programming guide \cite{CUDA_Programming_Guide}.

\subsection{Basic Implementation}
As discussed at the end of \cref{sec:apsm}, to access the filter estimate $f_{n}$ in \eqref{eqn:apsm}, we require access to the overall dictionary $\mathcal{D}_n:=\{\mathcal{D}_{\mathrm{G},n} \cup \mathcal{D}_{\mathrm{L}}\}$ and the coefficients $\gamma_i^{(n)}$ at time $n$. The new filter estimate $f_{n+1}$ is computed using projections on a sequence (sliding window) of $W$ convex sets constructed from $f_n$, the training sample received at time $n$, and $\mathcal{D}_n$. The Gaussian dictionary is then extended with the newly arriving training sample and, along with the new coefficients $\gamma_i^{(n+1)}$, it is used to store $f_{n+1}$.

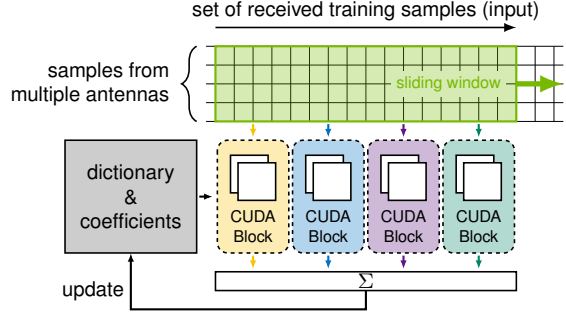
\begin{figure}[t] 
    %
    %
    \centering
    \scriptsize\centering\input{figures/tikz/apsm-training-cuda.tikz}
    \caption[training]{Training overview.
    The received vectors in the sliding window are processed by the \acs{CUDA} blocks in parallel, computing projections of $f_n$ onto the $W$ convex sets defined by the $W$ received vectors and the corresponding modulation symbols.
    The resulting estimate (bottom) is used to update the dictionary and the coefficients (left) then representing $f_{n+1}$.}
    \label{fig:cuda_training}
    %
    %
\end{figure}

Every training sample (the received vector \eqref{eqn:uplink_signal}) consists of samples from multiple antennas at time $n$ during the training phase. One \ac{CUDA} block can be used for each vector in the sliding window, with multiple threads inside the block computing the inner products $\left\langle f_n,\kappa(\mathbf{r}_j,\cdot)\right\rangle_{\mathcal{H}}$ in parallel. The final summation is performed in a reduction step, which can also be executed in parallel. As the new training samples arrives, the sliding window advances from $n$ to $n+1$ and the process is repeated. This implementation variant is depicted in \cref{fig:cuda_training}. Furthermore, as discussed in \cref{sec:acceleration_via_parallel_processing}, since both the inner products and the per-vector operations can be computed in parallel, we can use the GPU resources efficiently, as detailed later in this section.

\comment{%
For the creation of the dictionary and the coefficients, we use the pilots, as we know the symbols that correspond to the samples received.
As training samples are received, they are projected against a convex set, formed by the elements of $\mathcal{D}$ at time $n$.
These projections are used to create an estimation by means of \cref{eqn:apsm}, and the values of $\gamma_i^{(n)}$ adjusted by means of \Cref{eqn:beta}, respectively.

Because the inner products in \eqref{eqn:beta} involve independent kernel evaluations, these evaluations can be performed in parallel. As new training samples are received, the sliding window advances and the process is repeated. This implementation is depicted in \cref{fig:cuda_training}.
}%

\comment{%
The training stage is depicted in \cref{fig:cuda_training}.
It receives a set of training symbols and uses a sliding window to project them in parallel against the current set of symbols in the dictionary.
Since the expected symbol is known, an error figure can be computed and is used to update the values of weights and select or discard elements in the dictionary.
It is important to note that the dictionary contains only symbols received during this training phase.
As new samples are received during the training stage, the sliding window moves forward and the process is repeated.

The formulation described in the previous section can be realized with a series of projections of input vectors (samples) onto the vectors stored in a known dictionary. Such projections can be computed with the use of accumulations, multiply-add and exponential operations. Since they are mainly independent of each other and can be done in parallel, we aim for a massively parallel processor to accelerate this process.

The algorithm consists of two main stages, the online training focused on the generation of the dictionary and the weights from processing the received signals of known symbols, and the detection stage that uses the generated dictionary and weights to detect the symbols received during the transmission of data. Both stages are implemented in parallel for processing in real time in the \acs{GPU}. Both stages process the received samples asynchronously in batches, so a certain number of samples are collected and processed in the \ac{GPU} while the host systems keeps batching incoming samples from the \acp{SDR}.
}%

\begin{figure}[t] 
    %
    %
    \centering
    \scriptsize\centering\input{figures/tikz/apsm-detection-cuda.tikz}
    \caption[detection]{Detection overview.
    The top depicts the sequence of input vectors $\mathbf{r}_n$ for which we need to calculate $f_n(\mathbf{r}_n)$ as an estimate of $b_n$.
    Each input vector is processed by one $\acs{CUDA}$ block in parallel, and inside each $\acs{CUDA}$ block, the corresponding inner product is computed by multiple threads in parallel.
    The dictionary and the coefficients represent $f_n$.
    The result of each block is the approximation of the symbol $b_n$, which is stored in an array (bottom).}
    \label{fig:cuda_detection}
    %
    %
\end{figure}
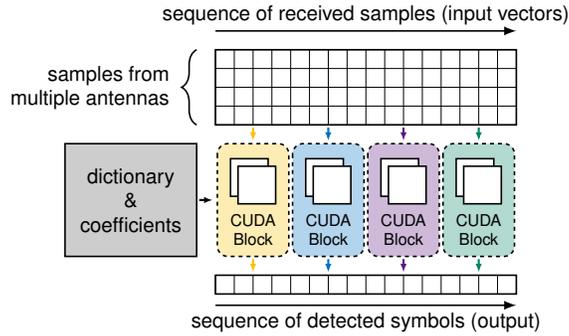

The detection stage computes $f_n(\mathbf{r}_n)$ (according to the formulas in \eqref{eqn:inner_product_definition} and \eqref{eqn:sum_inner_product}) as an estimate of $b_n$, where $f_n$ is the trained filter at detection time $n$ and $\mathbf{r}_n$ is the received vector. It does not modify $\mathcal{D}_n$ or $\gamma_i^{(n)}$, instead using them to access $f_{n}:=\sum_{i=1}^{n-1}\gamma^{(n)}_{i}\kappa(\mathbf{r}_i,\cdot)$.
Because $f_n(\mathbf{r}_n)$ is essentially the inner product $\left\langle f_n,\kappa(\mathbf{r}_n,\cdot)\right\rangle_{\mathcal{H}}$, the estimation can also be parallelized by processing each input vector in parallel, one per block. The required $n-1$ inner products are computed one per thread within the block. Finally, the results from threads in a block are summed to produce an estimation from the input vector as $f_{n}(\mathbf{r}_n):=\sum_{i=1}^{n-1}\gamma^{(n)}_{i}\kappa(\mathbf{r}_i,\mathbf{r}_n)$. This is depicted in \cref{fig:cuda_detection}.

\comment{%
and its job is to compute $f_n(\mathbf{r}_n)$ (as estimate of $b_n$), where $f_n$ is the trained filter at detection time $n$ and $\mathbf{r}_n$ is the received vector.
Because $f_n(\mathbf{r}_n)$ is nothing but the inner product $\left\langle f_n,\kappa(\mathbf{r}_n,\cdot)\right\rangle_{\mathcal{H},\signal{w}}$, the above-mentioned inner product procedure is repeated here.
For this, we use $f_{n}:=\sum_{i=1}^{n-1}\gamma^{(n)}_{i}\kappa(\mathbf{r}_i,\cdot)$, so we compute the inner product of the received vector against the dictionary elements and get the best estimate value as the sum of each weighted product.
As for further parallelization opportunities, notice that the processing of each $f_n(\mathbf{r}_n)$ during detection is independent, so these are computed in parallel across multiple \ac{CUDA} blocks.
Also since each $f_{n}(\mathbf{r}_n)$ is decomposed as $f_{n}(\mathbf{r}_n):=\sum_{i=1}^{n-1}\gamma^{(n)}_{i}\kappa(\mathbf{r}_i,\mathbf{r}_n)$, this sum is distributed across multiple \ac{CUDA} threads inside each block. 
Also, since the inner products are independent of each other, this can also be done in parallel inside each block and is therefore distributed across multiple \ac{CUDA} threads.
}%

We point out that further parallelization is possible as the detection of each desired user can be performed independently. The number of users that can be detected in parallel is therefore only limited by the \ac{GPU} resources available.

\comment{Furthermore, since each desired user can be detected independently, we can execute all procedures in parallel, as much as the \ac{GPU} resources allow it.}

\subsection{Optimization of Inner Products}
\label{sec:Optimization}
Both training and detection share the computation of the inner products $\left\langle f_n,\kappa(\mathbf{r}_n,\cdot)\right\rangle_{\mathcal{H}}$, which is the dominant operation of projection based algorithms.
Implemented as a \ac{CUDA} kernel to be executed on the \acs{GPU} (not to be confused with the kernel functions $\kappa$), it is the primary target for optimization with respect to the latency of the system.

Parameters like the number of threads per group, the number of groups per block, and the total amount of vectors cached in shared memory all influence the performance.
Device properties like type and speed grade of global memory, number of \acp{SM}, and amount of shared memory are important to consider.
The optimal set of parameters is found by sweeping the parameter space combining experimentation and guidance by profiling tools.

While optimizing the parameters determining the execution of the \ac{CUDA} kernel is part of  performance optimization, algorithm transformations for massively parallel architectures are required beforehand. In the following sections, we explore such principle  techniques that allow us to achieve the desired performance of a detection latency below \SI{1}{millisecond}.

\subsubsection{Multiple Sample Vectors per Block}
Our first technique aims at improving the \textit{occupancy} of the device, which is a measure of the utilization of the resources available within the \acs{GPU}.
Utilization, which is a function of several factors, can be maximized by allocating enough work to keep the device busy within the constraints of the hardware.
The most important constraints are the number of threads per \acs{SM}, the number of registers used per thread, and the amount of shared memory utilized by a \acs{CUDA} block.

To improve the occupancy of the \ac{CUDA} kernel, we describe the algorithm using Cooperative Groups (which are primitives for partitioning work inside a \ac{CUDA} block, see \cite{CUDA_Programming_Guide}) and assign the processing of an input vector $\mathbf{r}_n$ in \eqref{eqn:uplink_signal} to a single group.
The threads within this group are responsible for performing the parallel computations for a particular input vector, and the result is reduced (accumulated) at the end in parallel to obtain the estimated symbol.
We then assign multiple groups to one \ac{CUDA} block, with each group processing a different received vector within the block.
We can therefore use the number of groups in a block to maximize the occupancy of the \ac{SM}.

\subsubsection{Usage of Shared Memory}
\label{sec:SharedMemory}
Our second technique benefits from using the shared memory within the \acsp{SM} of a device.
As mentioned before, each block processes multiple input vectors and, for each input vector, computes inner products using the same dictionary entries.
Hence each entry of the dictionary may be loaded once and reused multiple times.
Shared memory is well suited for this because it is accessible to all groups within a block.
As a consequence global memory traffic is reduced by lowering the amount of data requested, reducing the time to load the data.

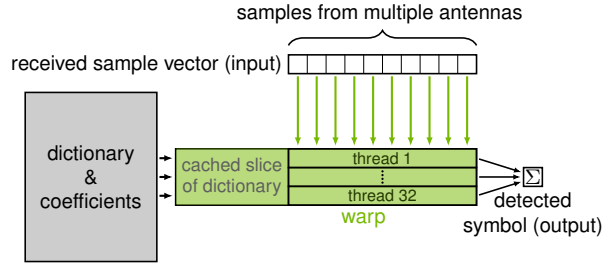
\begin{figure}[t] 
    \vspace*{-1.0\floatsep}
    \centering
    \scriptsize\centering\input{figures/tikz/shmem.tikz}
    \caption[shmem]{Shared Memory implementation.
    Each thread in the \ac{CUDA} block (center) computes the inner product using one dictionary entry by performing the required computations over the components of the vector $\mathbf{r}_n$ (top).
    Multiple elements of the dictionary (left) are cached in shared memory (center, left) in batches, allowing for the efficient processing of a subset of the dictionary.
    The final result is accumulated and stored as the approximated modulation symbol (right).}
    \label{fig:cuda_shmem}
    %
    %
\end{figure}

Every thread in the block reads the vector $\mathbf{r}_n$ from memory and then calculates the inner products with one entry of the cached shared slice of $\mathcal{D}_n$.
Once the inner products are calculated for this slice, a new slice of $\mathcal{D}_n$ is read into the cache and this process continues until the inner products for all members of $\mathcal{D}_n$ are computed.
This is depicted in \cref{fig:cuda_shmem} for the threads of one group; other groups (not depicted) in the same block operate over different input vectors but the same cached slice of the dictionary.

\subsubsection{Tuning and Hiding Latency}
While calculating the inner products, the \ac{CUDA} kernel spends most of its time accessing shared memory, and the computations that follow memory reads are faster than the memory  accesses, which can be verified by using a profiler tool.
Therefore, it is desirable to have a balance between memory accesses and computation, so neither will dominate (and possibly limit) the performance.

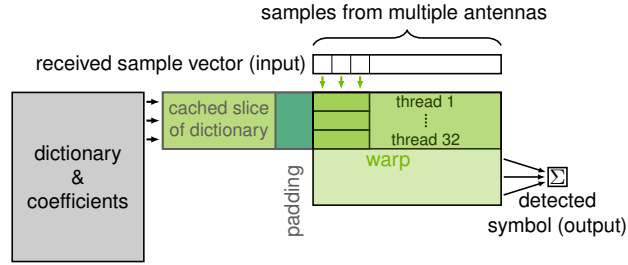
\begin{figure}[t] 
    \vspace*{-1.0\floatsep}
    \centering
    \scriptsize\centering\input{figures/tikz/balanced.tikz}
    \caption[balanced]{Balanced implementation.
    Threads in a \acs{CUDA} block process subsets of both the input vectors and the dictionary entries in contrast to \cref{fig:cuda_shmem}.
    Since only a subset of the vector $\mathbf{r}_n$ is processed, the intermediate results (before the entire $\mathbf{r}_n$ is processed) have to be temporarily stored. 
    The size of the cached dictionary elements is independent of the number of threads in the block (in contrast to \cref{fig:cuda_shmem}), and it is tuned to balance the time spent in computing and data load.}
    \label{fig:cuda_balanced}
    %
    %
\end{figure}

\comment{%
To balance the memory accesses with the computation of the inner products, it is possible to hide some of the latency from the memory accesses by computation in the \acp{SM}.
To achieve this, it is possible to increase the size of the cache being used to store the dictionary, therefore changing the distribution of load and compute times.
This helps because reads from global memory become more efficient, and because more computations can be performed with a bigger chunk of the dictionary available in faster memory.
In the process, the algorithm is also modified to process the vectors not as fixed lengths but as fixed chunks, which improves the load/compute pattern and allows the algorithm to address vectors of variable length while still use the same number of threads in the group.
The chunks then have to process all elements and save intermediate values to compute the inner products, but this amount of extra memory is not critical.
This procedure is depicted in \cref{fig:cuda_balanced}.
}%
\comment{%
Note that both the inner products (\ie, linear and the Gaussian kernel evaluations) require access to the received vector $\mathbf{r}_n$ and members of the dictionary $\mathcal{D}_n$.
In more detail, a naive implementation would read the vector $\mathbf{r}_n$ in its entirety from the memory and then calculate the inner products with the cached slice of $\mathcal{D}_n$.
Once the inner products are calculated with the cached slice, a new slice of $\mathcal{D}_n$ is read into the cache and this process continues until the inner products with all members of $\mathcal{D}_n$ are accomplished.
This process is shown in \cref{fig:cuda_shmem}.
An alternate way to achieve the above result faster, is to access only a part of the vector $\mathbf{r}_n$ (see \cref{fig:cuda_balanced}), and perform the computations (required for the inner products) with a larger chunk of dictionary (larger than the case in \cref{fig:cuda_shmem}) in the cached slice.
In this case, the intermediate results have to be stored until the entirety of $\mathbf{r}_n$ has been processed, but this extra memory usage is negligible.
This ``rearrangement'' of the inner product computations is shown in \cref{fig:cuda_balanced}, and it achieves a better balance between memory access and computations.
Note that the internal GPU mechanisms may also (where possible) overlap computations and memory reads if they are independent.
We noted that the above-mentioned rearrangement decreases the inner product computation times by half on average. 
}%

Both the inner products (\ie, the linear and the Gaussian kernel evaluations) require access to the received vector $\mathbf{r}_n$ and members of the dictionary $\mathcal{D}_n$.
The simple approach described in \cref{sec:SharedMemory} can be optimized by changing the order of the computations and increasing the amount of data in the cached slice (see \cref{fig:cuda_balanced}).
By fetching only a subset of $\mathbf{r}_n$ once and reusing it across a bigger section of the cache, without computing the full inner product in one go, the number of cache loads is reduced and the efficiency is increased.
Note that this comes at a relatively small cost of requiring intermediate storage in shared memory until the entire $\mathbf{r}_n$ is processed.
This ``rearrangement'' of the inner product computations is shown in \cref{fig:cuda_balanced}, and it achieves a better balance between memory access and computations, allowing for load and computation to overlap and decrease the overall inner product computation time by about half. 

%% file: figures/tikz/cuda-device-mapping.tikz
\begin{tikzpicture}[
		scale = 0.3,
		desc/.style={inner sep = 0, outer sep = 0}]

		\draw[thick, fill = nvtextcolor, fill opacity = 0.15]
			(0, -0.5) -- (19.5, -0.5) --
			(19.5, -0.5) .. controls ++(0.5, 0) and ++(0, 0) .. ++(0.5, -0.5) --
			(20, -1.5) .. controls ++(0, -0.5) and ++(0, 0) .. ++(0.5, -0.5) --
			(26.5, -2) .. controls ++(0.5, 0) and ++(0, 0) .. ++(0.5, 0.5) --
			(27, 10.5) .. controls ++(0, 0.5) and ++(0, 0) .. ++(-0.5, 0.5) --
			(0, 11) .. controls ++(-0.5, 0) and ++(0, 0) .. ++(-0.5, -0.5) --
			(-0.5, 0) .. controls ++(0, -0.5) and ++(0, 0) .. ++(0.5, -0.5) -- cycle;
		\node[font=\Large] at (23.5, -1.25) { GPU };

		\foreach \by in {0, 1}{
			\foreach \bx in {0, 1, 2.25}{
				\draw[thick, fill = white] (8.2*\bx, 6.5*\by) rectangle ++(8, 4);
				\foreach \x in {0, 1, ..., 7}{
					\foreach \y in {0, 1, ..., 3}{
						\draw (8.2*\bx+\x, 6.5*\by+\y) rectangle ++(1, 1);
					}
				}
			}
			\node at (17.33, 2 + 6.5*\by) {...};
		}

		\foreach \bx in {0, 1}{ 
			\node at (4+8.2*\bx, 5.5) {$\vdots$};
		}

		\draw[thick, HighlightColor, fill = HighlightColor, fill opacity = 0.5] (2.25*8.2,6.5) rectangle ++(1, 1);
		\node[HighlightColor, anchor = north west, inner sep = 0, outer sep = 0] at (2.25*8.2,6.5 - 0.25) { CUDA Core };

		\draw[thick, AmazonColor, fill = AmazonColor, fill opacity = 0.5] (2.25*8.2, 0) rectangle ++(8, 4);
		\node[AmazonColor, anchor = south, inner sep = 0, outer sep = 0 ] at (2.25*8.2 + 2.8, 4.25) { Streaming Multiprocessor (SM) };
\end{tikzpicture}

%% file: figures/tikz/apsm-training-cuda.tikz
\begin{tikzpicture}
	[
		scale = 0.5,
		every node/.style={font = \relsize{.5}\sffamily},
		thicker/.style={line width=.6pt},
		thinner/.style={line width=.2pt},
		shortened/.style={shorten <= 1, shorten >= 1},
	]

    \useasboundingbox (-6,1.5) rectangle (10,-7.5);
    
	\draw[thicker, -latex] (0, 0.5) -- ++(8, 0) node[pos = 0.5, above] { set of received training samples (input) };
	\foreach \i in {0, 1, ..., 18}
	{
		\draw[thinner] (.5*\i, 0) -- (.5*\i, -2);
	}
	\foreach \i in {0, 1, ..., 4}
	{
		\draw[thinner] (-0.25, -.5*\i) -- (9.25, -.5*\i);
	}
	\draw[thicker, decorate, decoration={brace, amplitude=.2cm}] (-0.5, -2) -- (-0.5, 0) node[pos = 0.5, anchor = east, outer sep = 8, align = right] { samples from\\multiple antennas };

	\draw[HighlightColor, thick, fill = HighlightColor, fill opacity = 0.3] (0, 0) rectangle ++(8, -2);
	\draw[HighlightColor, line width=0.07cm, -latex] (8, -1) -- ++(1.25, 0);
	\node[HighlightColor, fill=HighlightColor!30, anchor = south, scale = 0.8] at (6.2, -1.35) { sliding window };

	\foreach \c/\i in {NVFluorite/0, NVCPUBlue/1, NVAmethyst/2, NVEmerald/3}
	{
		\begin{scope}[shift={(2*\i, -2.5)}]
			\draw[shortened, \c, thick, -{Latex[scale=0.5]}] (1, 0.5) -- (1, 0);
			\draw[fill = \c, fill opacity = 0.3, rounded corners, thicker, dash pattern={on 1.5pt off 1.5pt}] (0.05, 0) rectangle ++(1.9, -3);
			\draw[thicker, fill = white] (0.4, -0.4) rectangle ++(1, -1);
			\draw[thicker, fill = white] (0.6, -0.6) rectangle ++(1, -1);
			\node[anchor = south, align = center, outer sep = 3, scale = 0.8] at (1, -3) { CUDA\\Block };
			\draw[shortened, \c, thick, -{Latex[scale=0.5]}] (1, -3) -- ++(0, -0.5);
		\end{scope}
	}

	\draw[thicker, fill=NVLightGray] (-4, -2.5) rectangle ++(3.5, -3) node[pos = 0.5, align = center] { dictionary\\ \&\\ coefficients };
	\draw[shortened, thick, -{Latex[scale=0.5]}] (-0.5, -4) -- ++(0.5, 0);

	\draw[thicker] (0, -6) rectangle (8, -6.5) node[pos = 0.5] {$\Sigma$};
	\draw[thick, -{Latex[scale=0.5]}] (4, -6.5) -- (4, -7.0) -- (-2.25, -7.0) -- (-2.25, -5.5);
	\node[anchor = east] at (-2.25, -6.5) { update };

\end{tikzpicture}

%% file: figures/tikz/apsm-detection-cuda.tikz
\begin{tikzpicture}
	[
		scale = 0.5,
		every node/.style={font = \relsize{.5}\sffamily},
		thicker/.style={line width=.6pt},
		thinner/.style={line width=.2pt},
		shortened/.style={shorten <= 1, shorten >= 1},
	]

    \useasboundingbox (-6,1.5) rectangle (10,-7.5);

	\draw[thicker, -latex] (0, 0.5) -- ++(8, 0) node[pos = 0.5, above] { sequence of received samples (input vectors) };
	\foreach \i in {0, 1, ..., 16}
	{
		\draw[thinner] (.5*\i, 0) -- (.5*\i, -2);
	}
	\foreach \i in {0, 1, ..., 4}
	{
		\draw[thinner] (0, -.5*\i) -- (8, -.5*\i);
	}
	\draw[thicker, decorate, decoration={brace, amplitude=.2cm}] (-0.5, -2) -- (-0.5, 0) node[pos = 0.5, anchor = east, outer sep = 8, align = right] { samples from\\multiple antennas };

	\draw[thicker] (0, 0) rectangle (8, -2);
	\foreach \c/\i in {NVFluorite/0, NVCPUBlue/1, NVAmethyst/2, NVEmerald/3}
	{
		\begin{scope}[shift={(2*\i, -2.5)}]
			\draw[shortened, \c, thick, -{Latex[scale=0.5]}] (1, 0.5) -- (1, 0);
			\draw[fill = \c, fill opacity = 0.3, rounded corners, thicker, dash pattern={on 1.5pt off 1.5pt}] (0.05, 0) rectangle ++(1.9, -3);
			\draw[thicker, fill = white] (0.4, -0.4) rectangle ++(1, -1);
			\draw[thicker, fill = white] (0.6, -0.6) rectangle ++(1, -1);
			\node[anchor = south, align = center, outer sep = 3, scale = 0.8] at (1, -3) { CUDA\\Block };
			\draw[shortened, \c, thick, -{Latex[scale=0.5]}] (1, -3) -- ++(0, -0.5);
		\end{scope}
	}

	\draw[thicker, fill=NVLightGray] (-4, -2.5) rectangle ++(3.5, -3) node[pos = 0.5, align = center] { dictionary\\ \&\\ coefficients };
	\draw[shortened, thick, -{Latex[scale=0.5]}] (-0.5, -4) -- ++(0.5, 0);

	\foreach \i in {0, 1, ..., 16}
	{
		\draw[thinner] (.5*\i, -6) -- ++(0, -.5);
	}
	\draw[thicker] (0, -6) rectangle (8, -6.5);
	\draw[thicker, -latex] (0, -6.8) -- ++(8, 0) node[pos = 0.5, below] { sequence of detected symbols (output) };

\end{tikzpicture}

%% file: figures/tikz/shmem.tikz
\begin{tikzpicture}
	[
		scale = 0.5,
		every node/.style={font = \relsize{.5}\sffamily},
		thicker/.style={line width=.6pt},
		thinner/.style={line width=.2pt},
		shortened/.style={shorten <= 1, shorten >= 1},
	]
	
	\def\offset{2}
	
	\def\padding{0}
	\def\csize{1}
	
	\useasboundingbox (-5,-9.25) rectangle ++(17.5,7.5);
	
	\draw[thicker] (3 + \padding, -3.5) rectangle (8 + \padding, -4);
		\node[anchor = east] at (3 + \padding, -3.75) { received sample vector (input) };
	\foreach \i in {1, 2, ..., 10}
	{
		\draw[thinner] (.5*\i + 3 + \padding, -3.5) -- ++(0, -.5);
		\draw[HighlightColor, shortened, thick, -{Latex[scale=0.5]}] (3 + \padding + .5*\i-.25, -4) -- ++(0, -.5 - 1.5*(\offset-1););
	}
	\draw[thicker, decorate, decoration={brace, amplitude=.2cm}] (3 + \padding, -3.25) -- (8 + \padding, -3.25)
		node[pos = 0.5, anchor = south, outer sep = 6, align = right] { samples from multiple antennas };

	\draw[thicker, fill=NVLightGray] (-4, -4.5) rectangle ++(3.5, -4.5) node[pos = 0.5, align = center] { dictionary\\ \&\\ coefficients };

	\begin{scope}[shift={(0, -1.5*\offset+1.5)}]

		\foreach \i in {0, 1, 2}
		{
			\draw[shortened, thick, -{Latex[scale=0.5]}] (-0.5, -4.75-.5*\i) -- ++(0.5, 0);
		}
		
		\draw[thicker, fill = HighlightColor, fill opacity = 0.3] (0, -4.5) rectangle (8 + \padding, -6);
		\draw[thicker] (3 + \padding, -4.5) -- ++(0, -1.5);
		
		\fill[HighlightColor, fill opacity = 0.3] (0, -4.5) rectangle (3, -6)
			node[pos = 0.5, align = center, opacity = 1, nvtextcolor, scale = 0.9] { cached slice\\of dictionary };
		\if\padding0\else
			\draw[nvtextcolor, thicker, fill = NVEmerald, fill opacity = 0.6] (3, -4.5) rectangle ++(\padding, -1.5*\csize)
				node[anchor=east, opacity = 1, nvtextcolor, rotate = 90, opacity = 1] at (3.5,-6.25) { padding };
		\fi

		\foreach \i in {0, 1, 2}
		{
			\draw[thicker, -{Latex[scale=0.5]}, shorten >= 6, shorten <= 1] (8.0 + \padding, -4.75-.5*\i) -- (9.6 + \padding, -4.75-.5*1);
		}
		\draw[thicker] (9.25 + \padding, -5.0) rectangle ++(.5, -.5) node[pos = 0.5] { $\Sigma$ };
		\node[anchor = north, align = center] at (9.5 + \padding, -5.5) { detected\\symbol (output) };
		
		\foreach \i/\v in {0/thread 1, 1/\ , 2/thread 32}
		{
			\node[scale=0.8] at (5.5 + \padding, -4.75-.5*\i) { \v };
			\draw[thicker, fill = HighlightColor, fill opacity = 0.3] (3 + \padding, -4.5-.5*\i) rectangle ++(5, -.5);
		}
		\foreach \i in {0, 1, ..., 3}
		{
			\fill (5.5 + \padding, -5.1-.1*\i) circle (0.03);
		}
		\node[HighlightColor, anchor = north] at (5 + 1*\padding, -4.5 - 1.5*\csize) { warp };
		
	\end{scope}
	
\end{tikzpicture}

%% file: figures/tikz/balanced.tikz
\begin{tikzpicture}
	[
		scale = 0.5,
		every node/.style={font = \relsize{.5}\sffamily},
		thicker/.style={line width=.6pt},
		thinner/.style={line width=.2pt},
		shortened/.style={shorten <= 1, shorten >= 1},
	]
	
	\def\offset{1}
	
	\def\padding{1}
	\def\csize{1}
	
	\useasboundingbox (-5,-9.25) rectangle ++(17.5,7.5);
	
	\draw[thicker] (3 + \padding, -3.5) rectangle (8 + \padding, -4);
		\node[anchor = east] at (3 + \padding, -3.75) { received sample vector (input) };
	\foreach \i in {1, 2, ..., 3}
	{
		\draw[thinner] (.5*\i + 3 + \padding, -3.5) -- ++(0, -.5);
		\draw[HighlightColor, shortened, thick, -{Latex[scale=0.5]}] (3 + \padding + .5*\i-.25, -4) -- ++(0, -.5 - 1.5*(\offset-1););
	}
	\draw[thicker, decorate, decoration={brace, amplitude=.2cm}] (3 + \padding, -3.25) -- (8 + \padding, -3.25)
		node[pos = 0.5, anchor = south, outer sep = 6, align = right] { samples from multiple antennas };

	\draw[thicker, fill=NVLightGray] (-4, -4.5) rectangle ++(3.5, -4.5) node[pos = 0.5, align = center] { dictionary\\ \&\\ coefficients };

	\begin{scope}[shift={(0, -1.5*\offset+1.5)}]

		\foreach \i in {0, 1, 2}
		{
			\draw[shortened, thick, -{Latex[scale=0.5]}] (-0.5, -4.75-.5*\i) -- ++(0.5, 0);
		}
		
		\draw[thicker, fill = HighlightColor, fill opacity = 0.3] (0, -4.5) rectangle (8 + \padding, -6);
		\draw[thicker] (3 + \padding, -4.5) -- ++(0, -1.5);
		
		\draw[thicker, nvtextcolor, fill = HighlightColor, fill opacity = 0.3] (0, -4.5) rectangle ++(3, -1.5*\csize)
			node[pos = 0.5, align = center, opacity = 1, nvtextcolor, scale = 0.9] { cached slice\\of dictionary };
		\if\padding0\else
			\draw[nvtextcolor, thicker, fill = NVEmerald, fill opacity = 0.6] (3, -4.5) rectangle ++(\padding, -1.5*\csize)
				node[anchor=east, opacity = 1, nvtextcolor, rotate = 90, opacity = 1] at (3.5,-6.25) { padding };
		\fi

		\foreach \i in {0, 1, 2}
		{
			\draw[thicker, -{Latex[scale=0.5]}, shorten >= 6, shorten <= 1] (8.0 + \padding, -4.75-.5*\i -1.5*\csize) -- (9.6 + \padding, -4.75-.5*1 - 1.5*\csize);
		}
		\draw[thicker] (9.25 + \padding, -5.0-1.5*\csize) rectangle ++(.5, -.5) node[pos = 0.5] { $\Sigma$ };
		\node[anchor = north, align = center] at (9.5 + \padding, -5.5-1.5*\csize) { detected\\symbol (output) };
		
		\foreach \i/\v in {0/thread 1, 1/\ , 2/thread 32}
		{
			\node[scale=0.8] at (5.5 + 0.5 + \padding, -4.75-.5*\i) { \v };
			\draw[thicker, fill = HighlightColor, fill opacity = 0.3] (3 + \padding, -4.5-.5*\i) rectangle ++(1.5, -.5);
		}
		\foreach \i in {0, 1, ..., 3}
		{
			\fill (5.5 + 0.5 + \padding, -5.1-.1*\i) circle (0.03);
		}
		\draw[thicker, fill = HighlightColor, fill opacity = 0.3] (3 + 1*\padding, -4.5) rectangle ++(5, -3);
		\node[HighlightColor, anchor = north] at (5 + 1*\padding, -4.5 - 1.5*\csize) { warp };
		
	\end{scope}
	
\end{tikzpicture}

%% file: sections/IV_lab_setup.tex

\section{Experimental Setup}
\label{sec:SystemSetup}
\begin{figure}[t] 
    %
    %
	\centering
    \begin{subfigure}[b]{0.48\linewidth}
        \centering
        \whencolumns{\fbox{\includegraphics[height=12em]{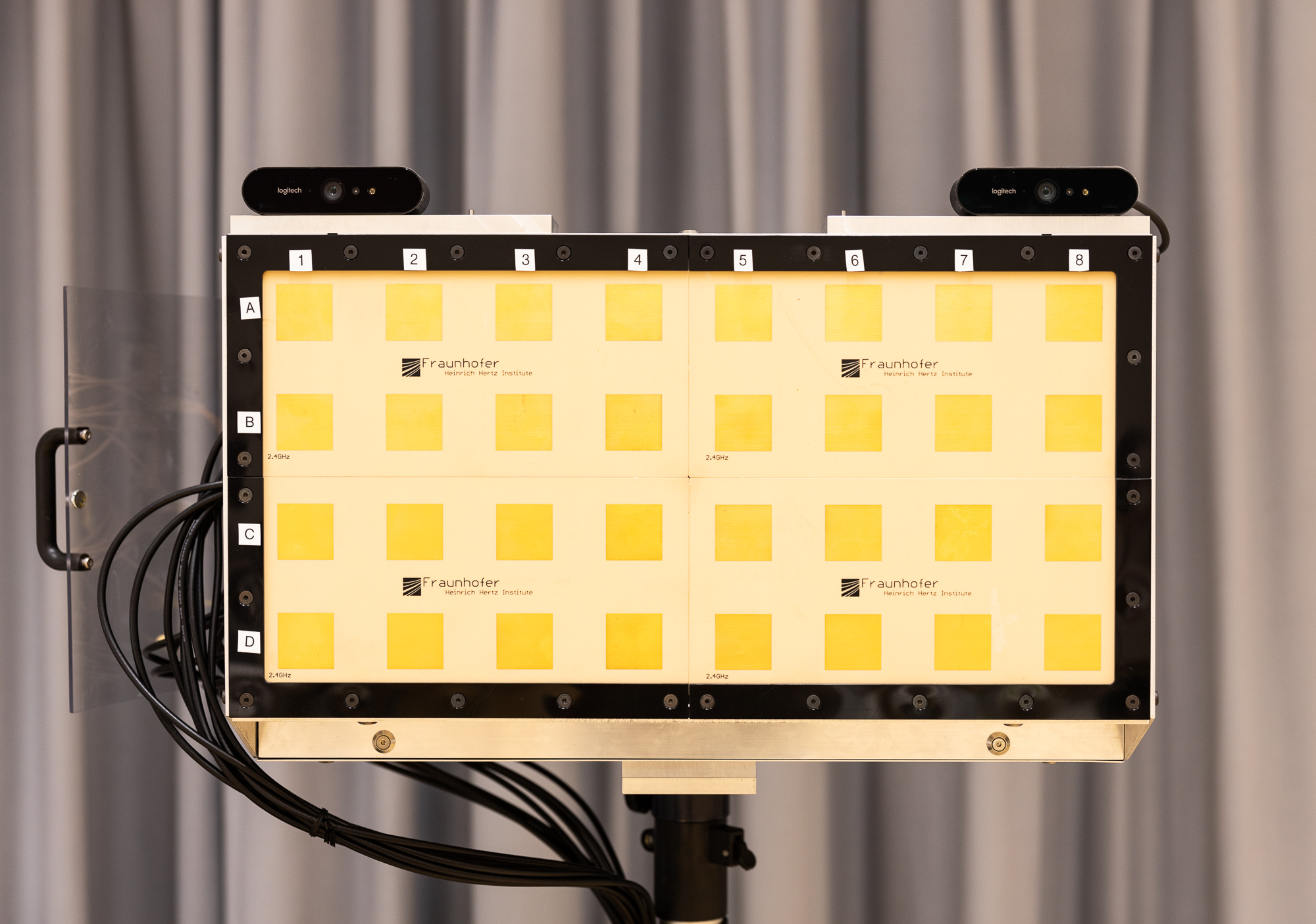}}}{\fbox{\includegraphics[height=8em]{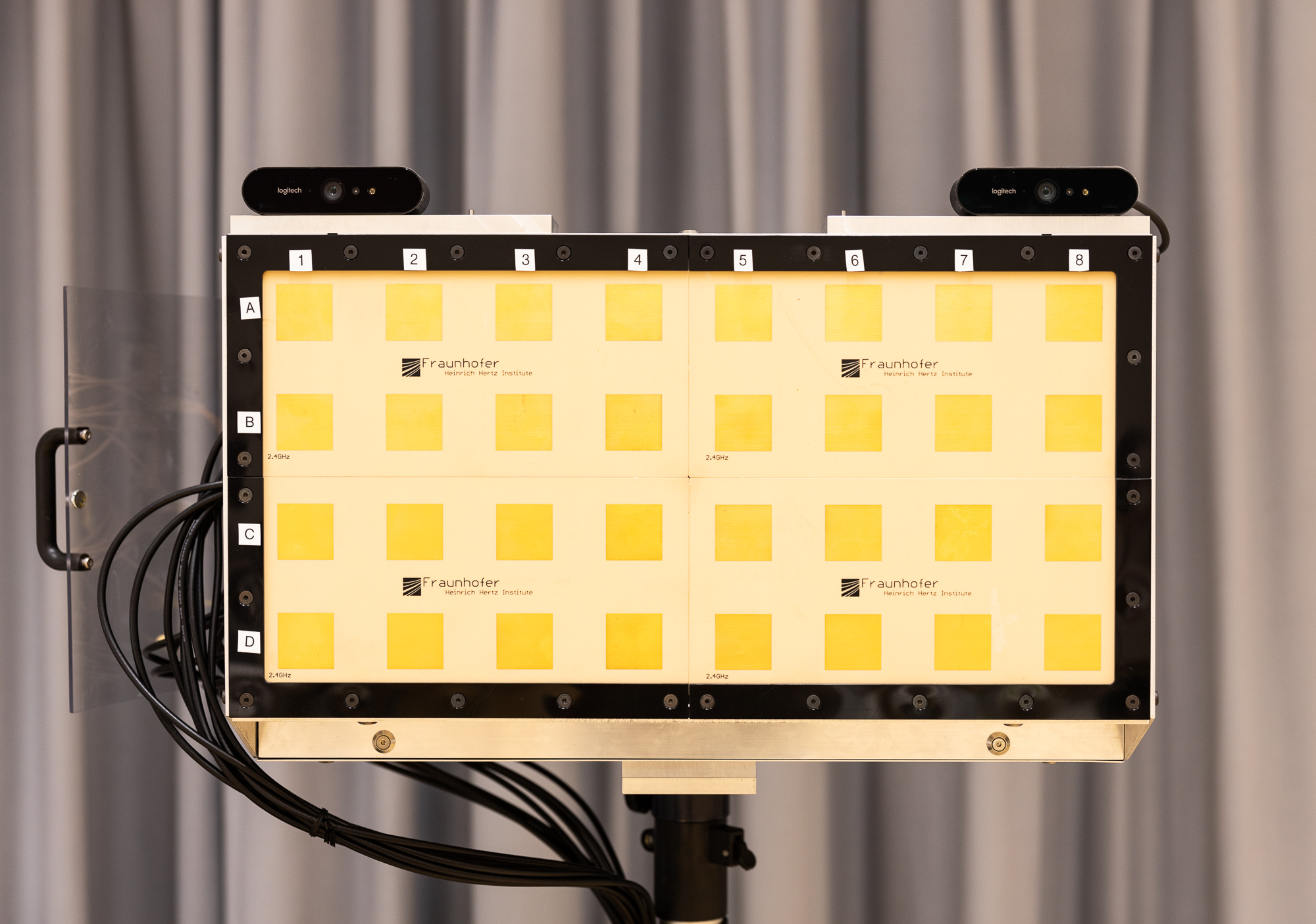}}}%
        \caption{Uniform Planar Array}
        \label{fig:NOMAplanar}
    \end{subfigure}
    \hfill
    \begin{subfigure}[b]{0.49\linewidth}
        \centering
        \whencolumns{\fbox{\includegraphics[height=12em]{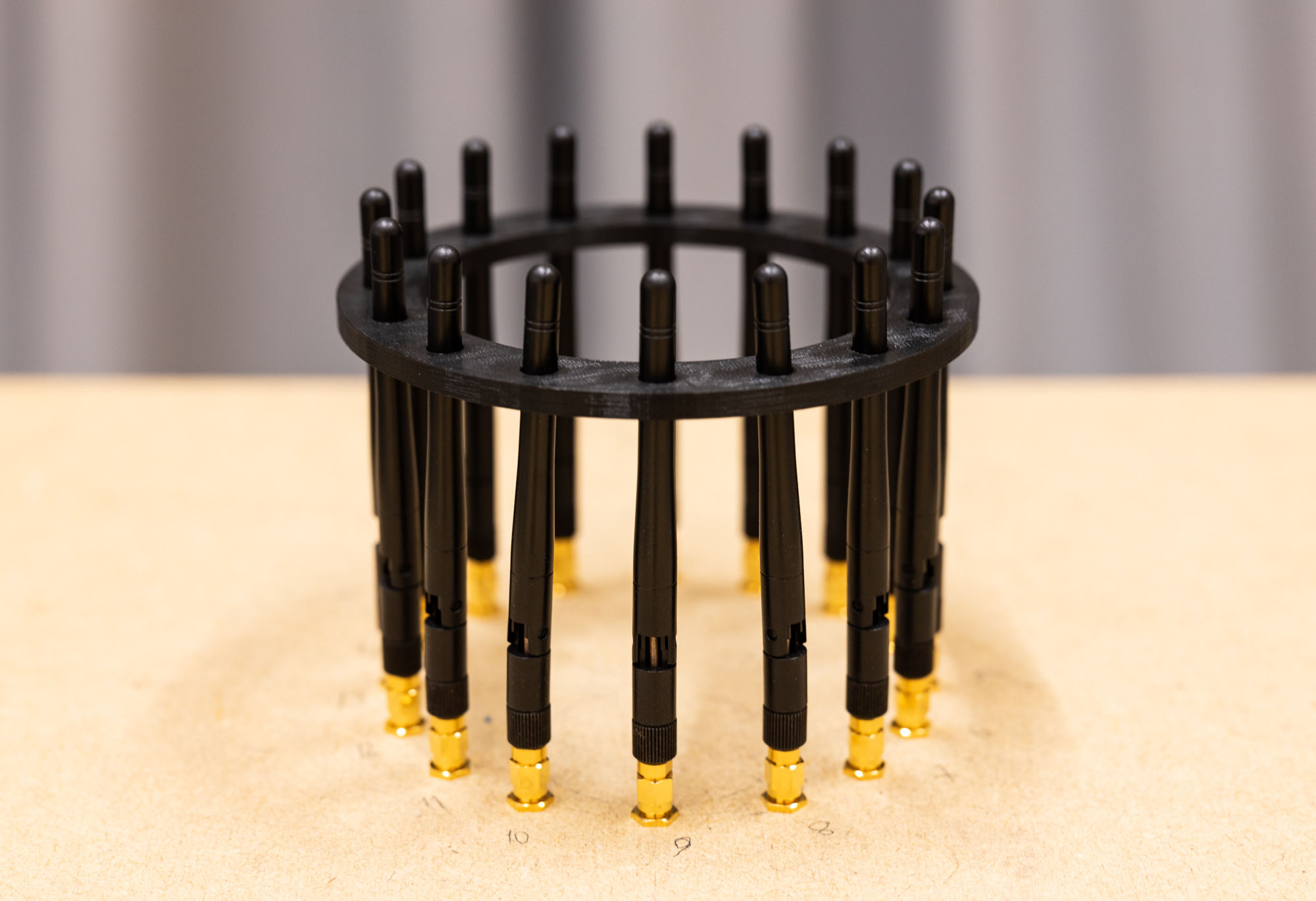}}}{\fbox{\includegraphics[height=8em]{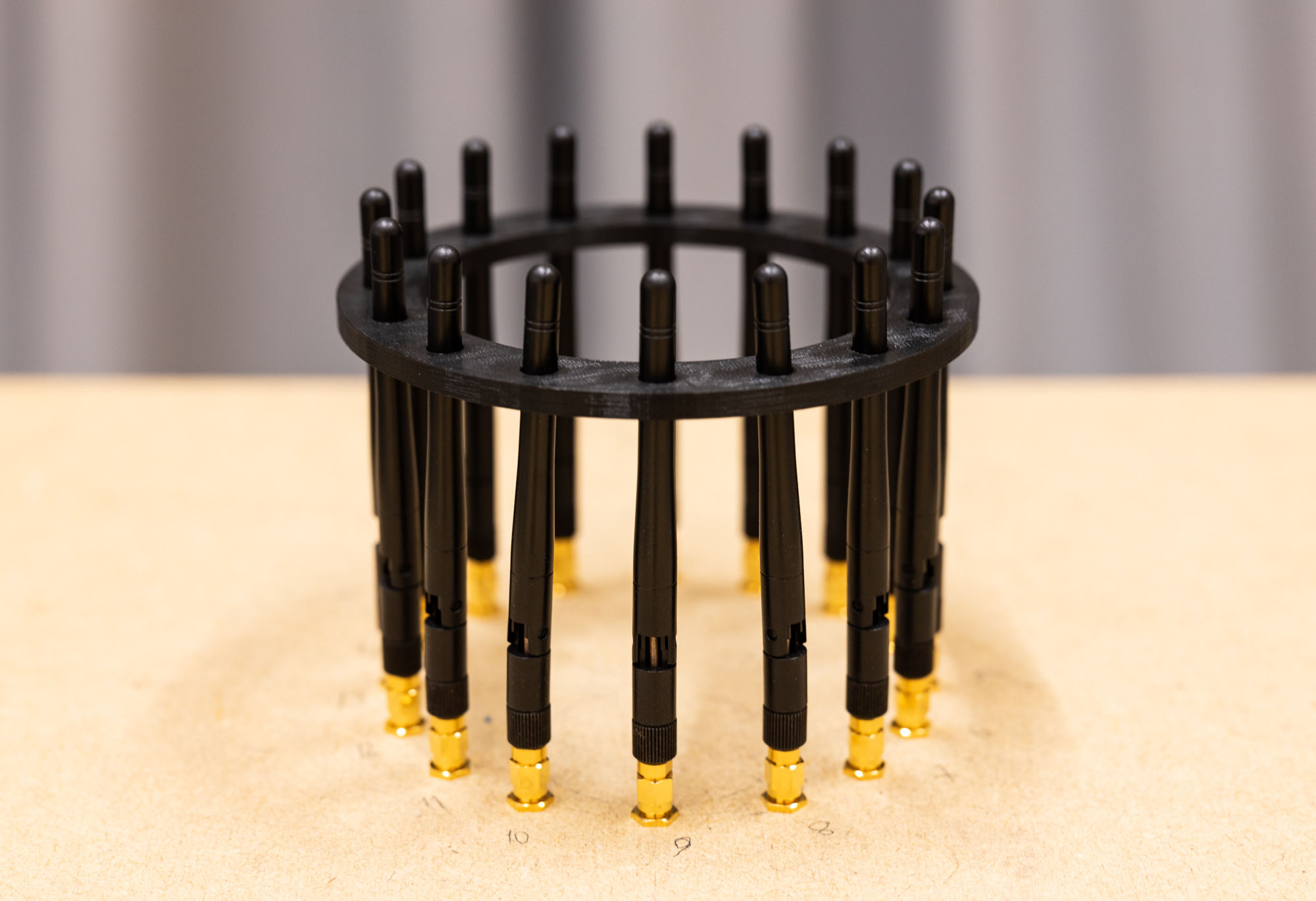}}}%
        \caption{Uniform Circular Array}
        \label{fig:NOMAcircular}
    \end{subfigure}
    \caption[NOMAAntennas]{Antenna array configurations.}
	\label{fig:NOMAAntennas}
	%
    %
\end{figure}

This section describes the hardware components of our real-time multiuser detection setup. 
The transmitter, receiver, and signal processing equipment are shown in \cref{fig:NOMAsetup}.
Except for the \ac{UPA} shown in \cref{fig:NOMAAntennas}, the components are \ac{COTS} devices.
\comment{%
\begin{figure*}[t] 
    %
    %
	\centering
	\begin{subfigure}[b]{0.14\linewidth}
        \centering
        \fbox{\includegraphics[height=8em]{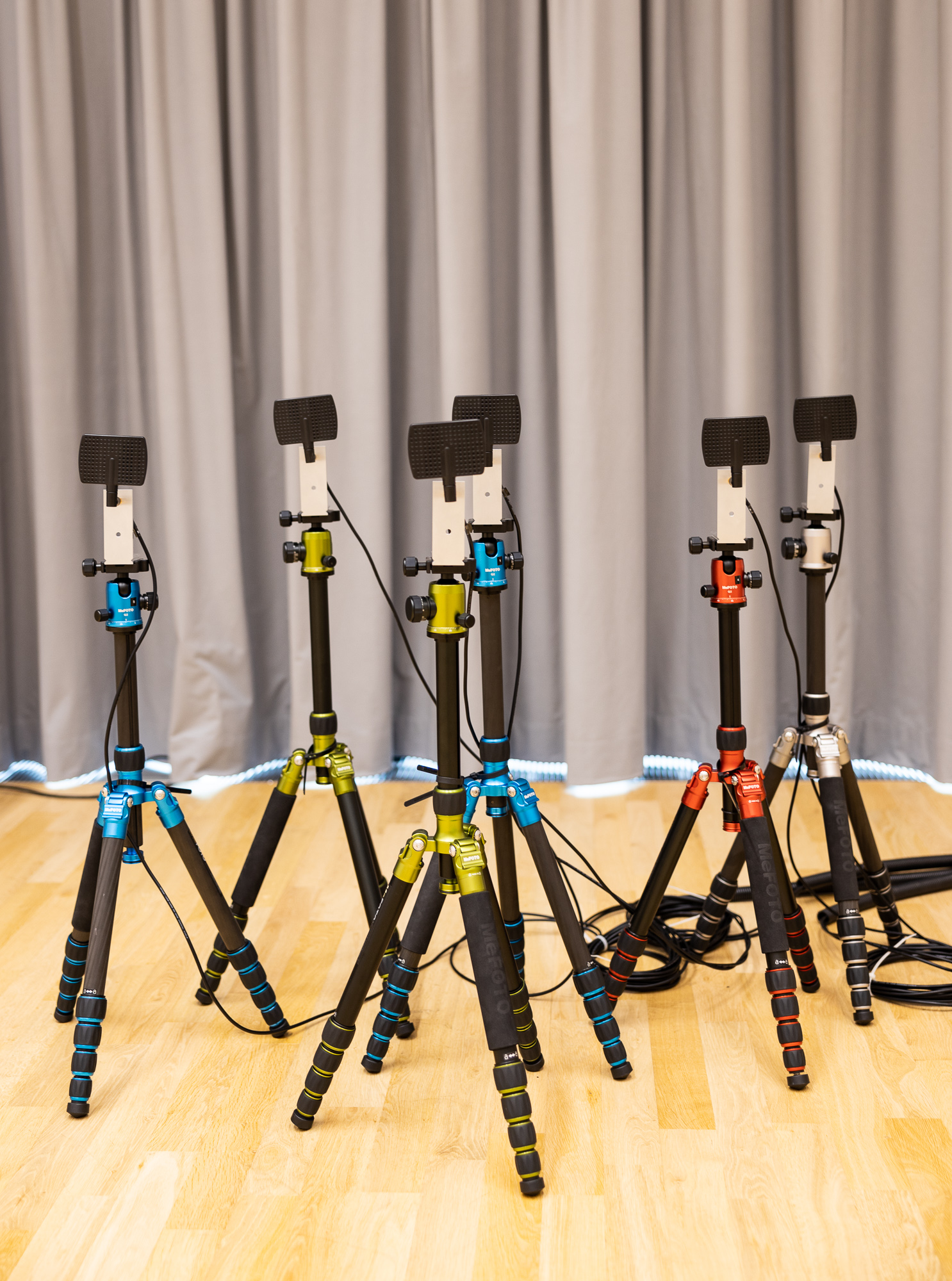}} 
        \caption{Users}
        \label{fig:NOMAusers}
    \end{subfigure}
    \hfill
    \begin{subfigure}[b]{0.28\linewidth}
        \centering
        \fbox{\includegraphics[height=8em]{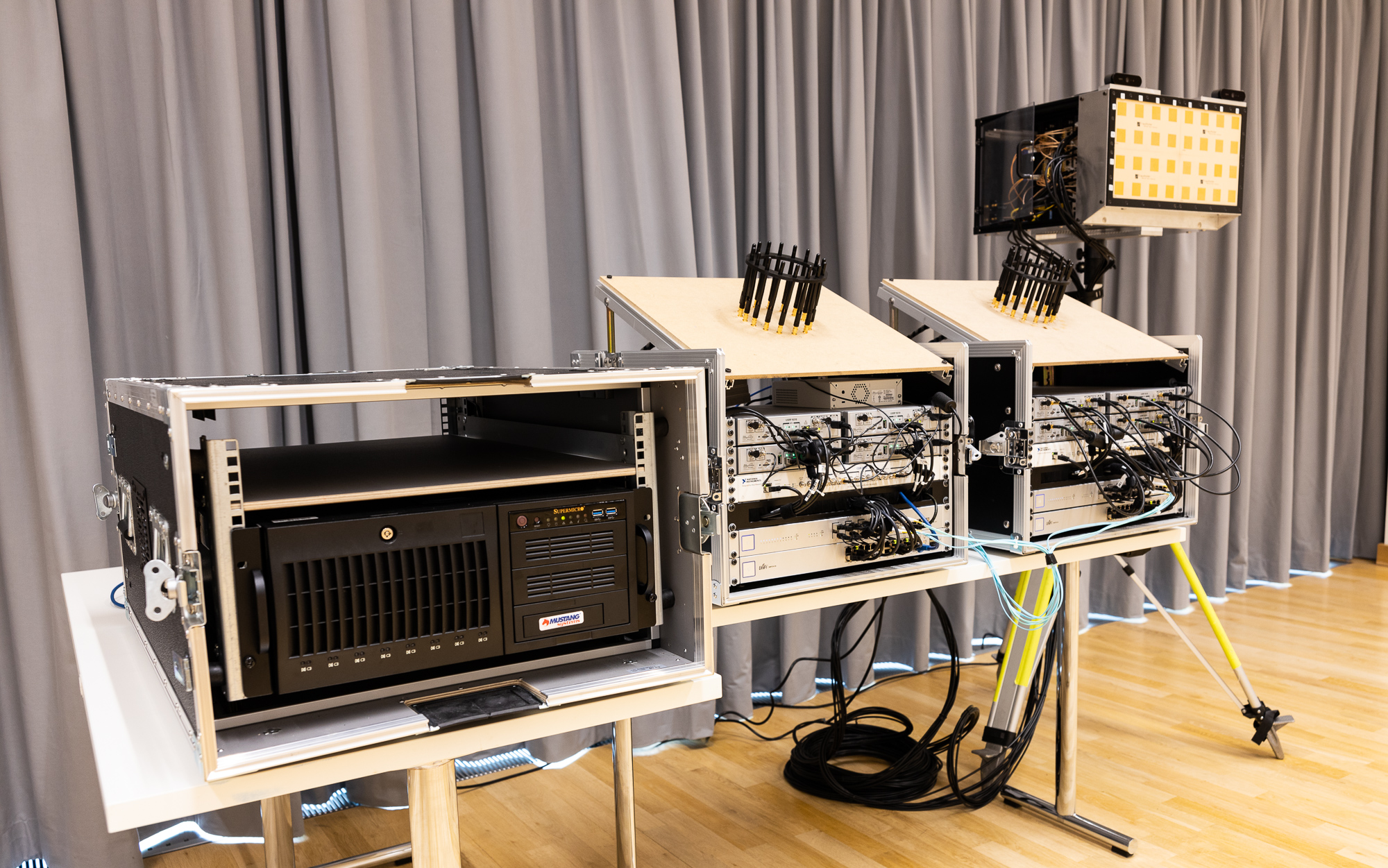}}
        \caption{Base station}
        \label{fig:NOMAbase}
    \end{subfigure}
    \hfill
    \begin{subfigure}[b]{0.25\linewidth}
        \centering
        \fbox{\includegraphics[height=8em]{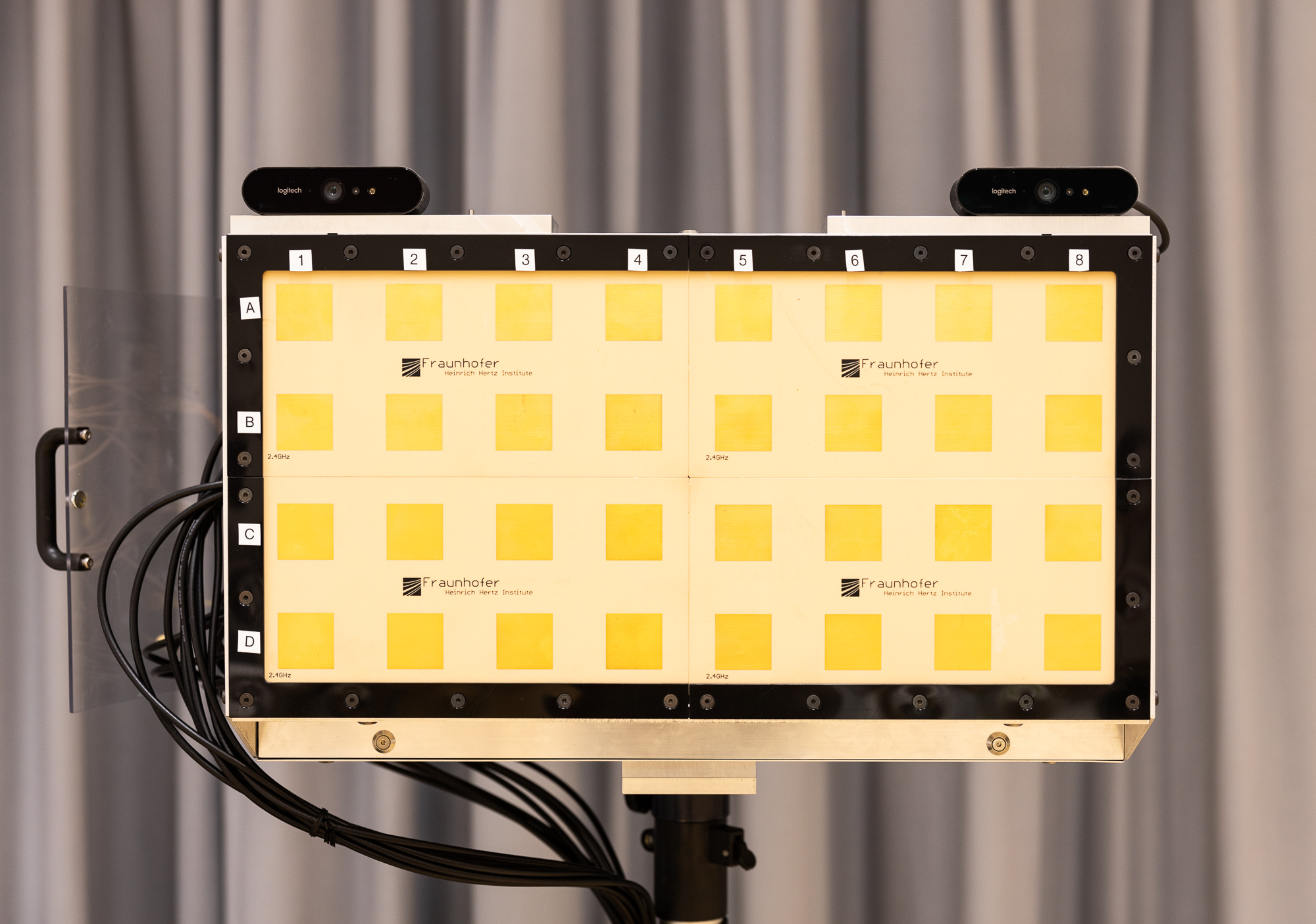}} 
        \caption{Planar Array}
        \label{fig:NOMAplanar}
    \end{subfigure}
    \hfill
    \begin{subfigure}[b]{0.26\linewidth}
        \centering
        \fbox{\includegraphics[height=8em]{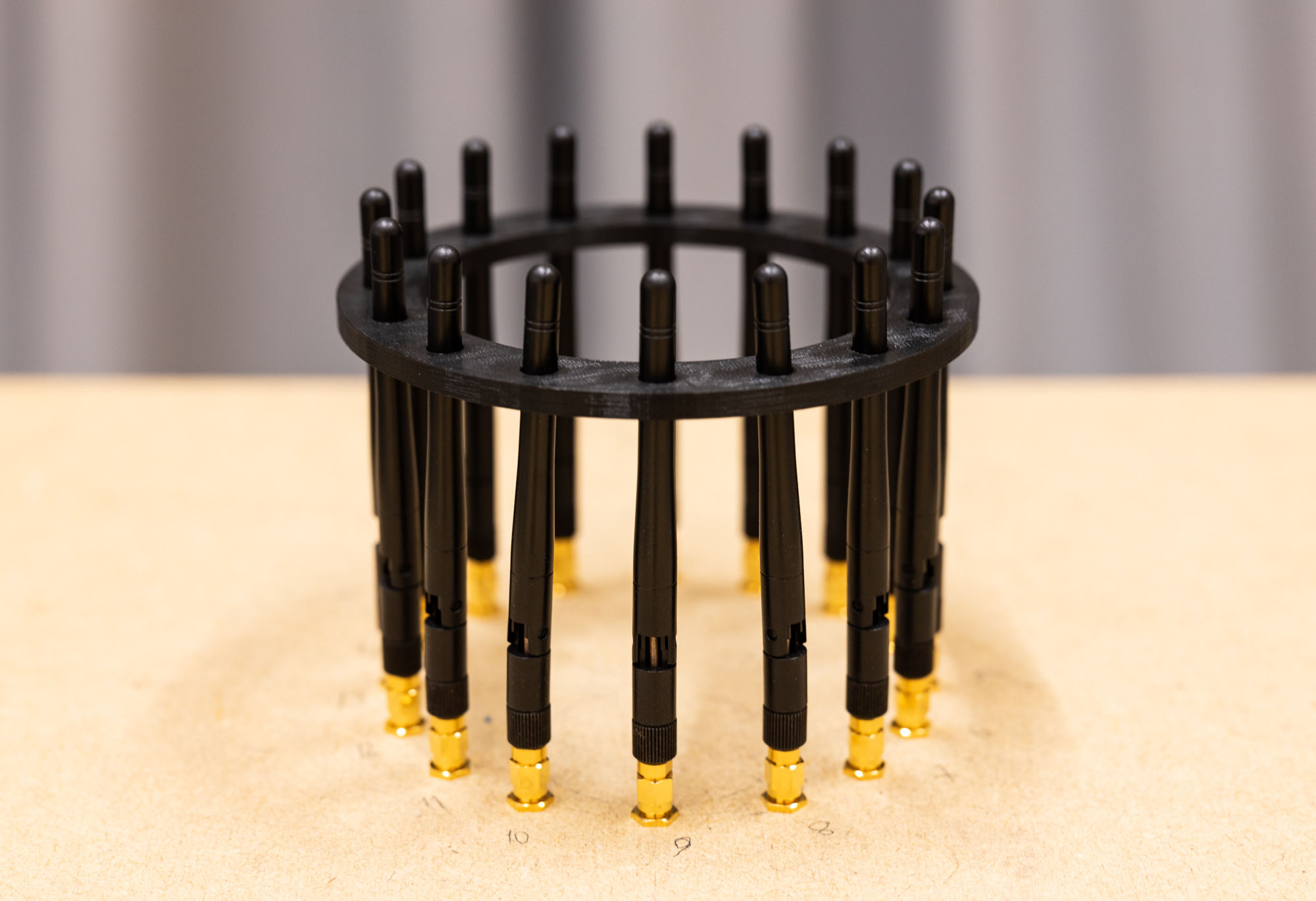}}
        \caption{Circular Array}
        \label{fig:NOMAcircular}
    \end{subfigure}
    \caption[NOMAsetup]{Real-time machine learning based multiuser detection setup.}
	\label{fig:NOMAsetup}
	%
    %
\end{figure*}
}%

\subsection{Signal Processing}
For development and testing, we use different locals servers and also several \ac{HPC} units that are equipped with multiple \ac{CPU} cores and at least one \ac{GPU}. 
Our local server is shown in \cref{fig:NOMAbase}, this server is equipped with a Xeon \mbox{W-3245} \ac{CPU} and two RTX~2080~Ti consumer \acp{GPU}.
This server also handles the signal processing part and the data transfer from and to the \acp{SDR}.

\subsection{Radio Access}
The basestation is composed of four Ettus \acs{USRP} N310 \acp{SDR}. 
Furthermore, we use a single \ac{NI} OctoClock for a \ac{GPS} disciplined clock and timing source for our \acp{SDR}.
With this setup, all four \acp{SDR}, each equipped with four ports on the \ac{Tx} and \ac{Rx} path, behave like a single \ac{SDR} system with up to sixteen synchronized physical antenna ports for both the \ac{Rx} and \ac{Tx} paths.

\begin{table}[t] 
	%
	%
	\centering
	    \whencolumns{\def\tablewidth {0.60\linewidth}}{\def\tablewidth {0.97\linewidth}}
        \begin{tabularx}{\tablewidth}{@{}*{1}{|R}*{2}{|c}|@{}}
			\hhline{~|-|-|} 
			\multicolumn{1}{r|}{}                                             & {\cellcolor{NVLightGray}}Value
			                                                                  & {\cellcolor{NVLightGray}}Comment                               \\ \hhline{*{3}{|-}|}
			{\cellcolor{NVLightGray}}Maximum system bandwidth                 & \SI{30.72}{\mega\hertz} &                                      \\ \hhline{*{3}{|-}|}
			{\cellcolor{NVLightGray}}\acs{UCA} radius                         & \SI{6.5}{\centi\metre}  &                                      \\ \hhline{*{3}{|-}|}
			{\cellcolor{NVLightGray}}\acs{UPA} patch resonant frequency       & \SI{2.442}{\giga\hertz} & $\lambda = \SI{12.28}{\centi\metre}$ \\ \hhline{*{3}{|-}|}
			{\cellcolor{NVLightGray}}\acs{UPA} patch element spacing          & \SI{6.14}{\centi\metre} & $\nicefrac{\lambda}{2}$              \\ \hhline{*{3}{|-}|}
			{\cellcolor{NVLightGray}}\acs{UPA} patch element length and width & \SI{3.07}{\centi\metre} & $\nicefrac{\lambda}{4}$              \\ \hhline{*{3}{|-}|}
			{\cellcolor{NVLightGray}}Used polarization                        & vertical                &                                      \\ \hhline{*{3}{|-}|}
			{\cellcolor{NVLightGray}}Number of \acs{Tx} \acs{SDR} modules     & 1                       & NAMC-SDR                             \\ \hhline{*{3}{|-}|}
			{\cellcolor{NVLightGray}}Number of \acs{Rx} \acs{SDR} modules     & 4                       & \acs{USRP} N310                      \\ \hhline{*{3}{|-}|}
			{\cellcolor{NVLightGray}}Number of \acs{Rx} antenna ports         & 16                      &                                      \\ \hhline{*{3}{|-}|}
		\end{tabularx}
	\caption{\acs{NOMA} system parameters.}
	\label{tab:NOMAparams}
	%
	%
\end{table}
A subset of our system parameters is shown in \cref{tab:NOMAparams} and is also given in the following \cref{sec:Res}.
For over the air signal transmission, we use an \ac{OFDM} system mode with a sampling rate of \SI{30.72}{\mega\hertz}.
Note that we can switch between radio signals from our \ac{UPA} shown in \cref{fig:NOMAplanar} and our \ac{UCA} shown in \cref{fig:NOMAcircular}.
The receive antenna arrays and user equipment (the transmitters) are described below. 

\begin{figure}[t] 
    %
    %
	\centering
	\hfill
    \begin{subfigure}[b]{0.56\linewidth}
        \centering
        \whencolumns{\fbox{\includegraphics[height=12em]{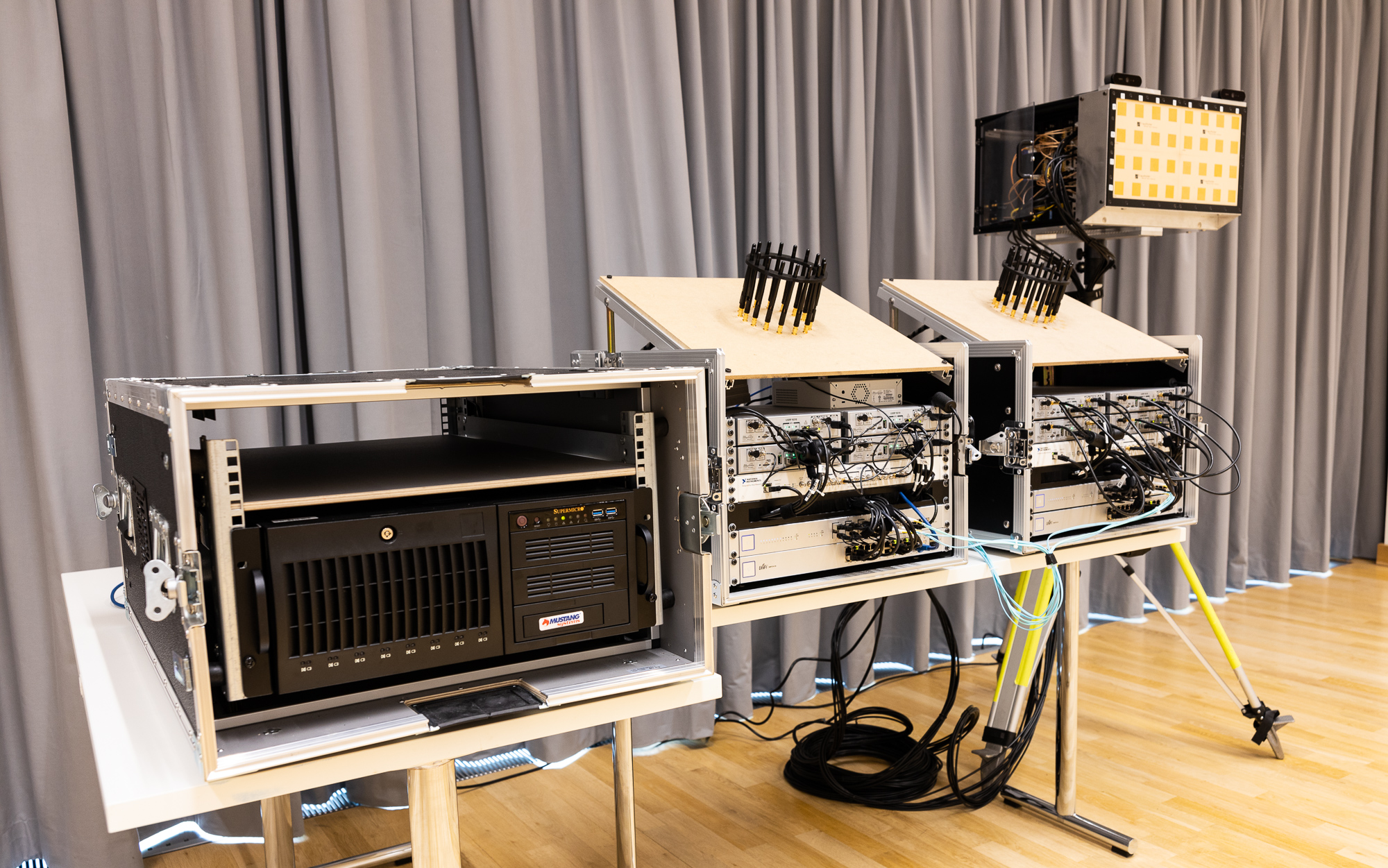}}}{\fbox{\includegraphics[height=8em]{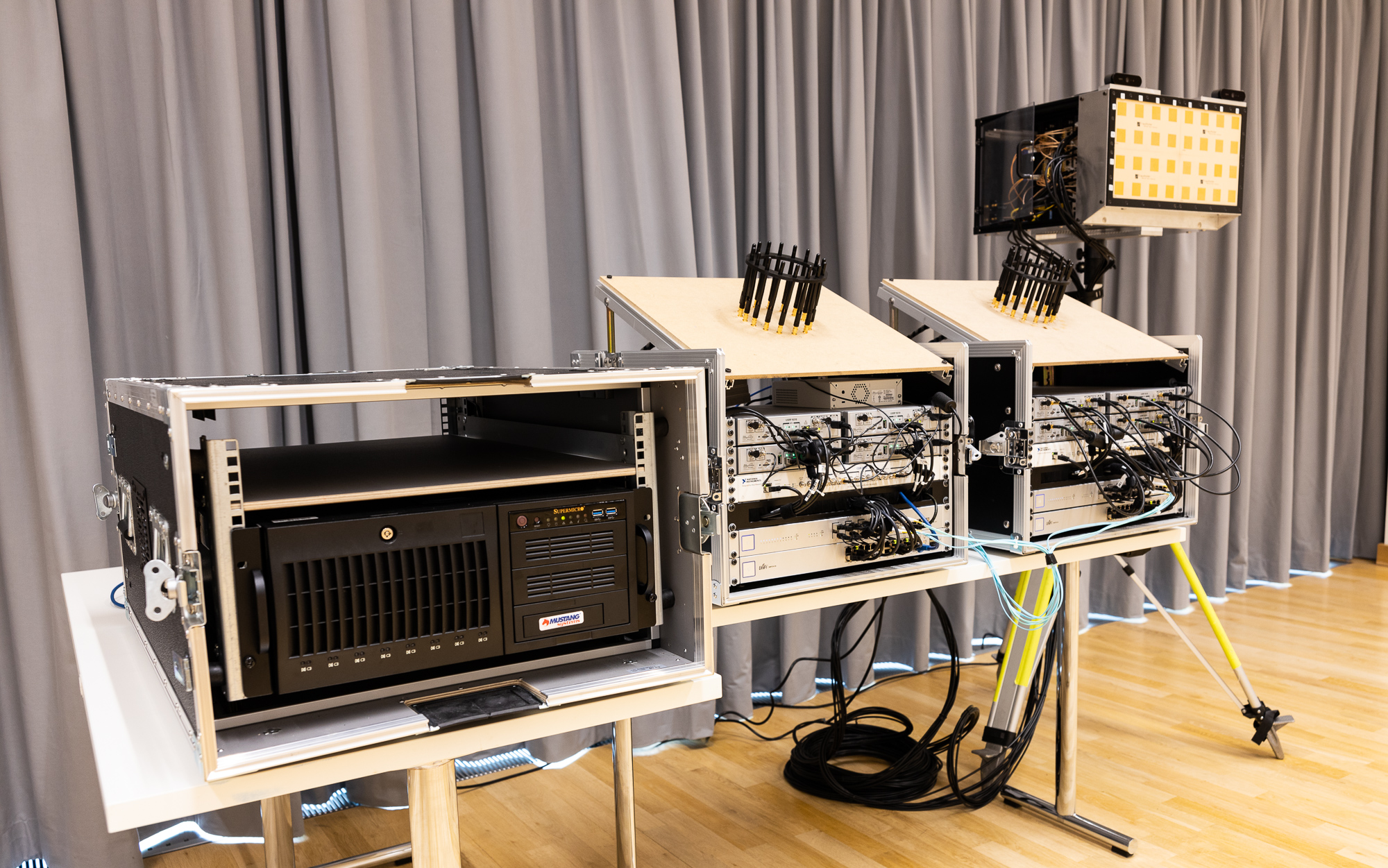}}}%
        \caption{Base station (\acl{Rx})}
        \label{fig:NOMAbase}
    \end{subfigure}
    \hfill
	\begin{subfigure}[b]{0.41\linewidth}
        \centering
        \whencolumns{\fbox{\includegraphics[height=12em]{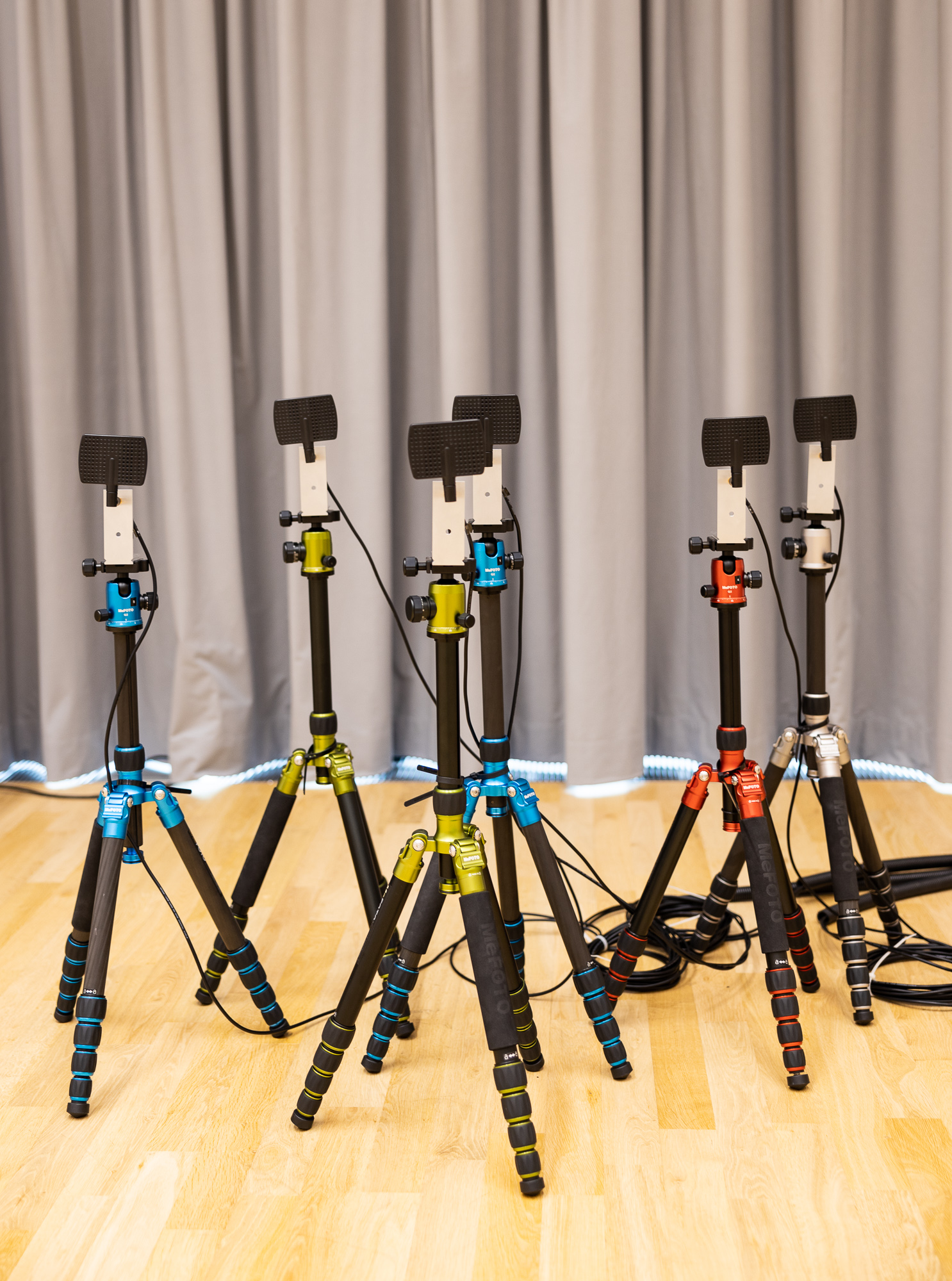}}}{\fbox{\includegraphics[height=8em]{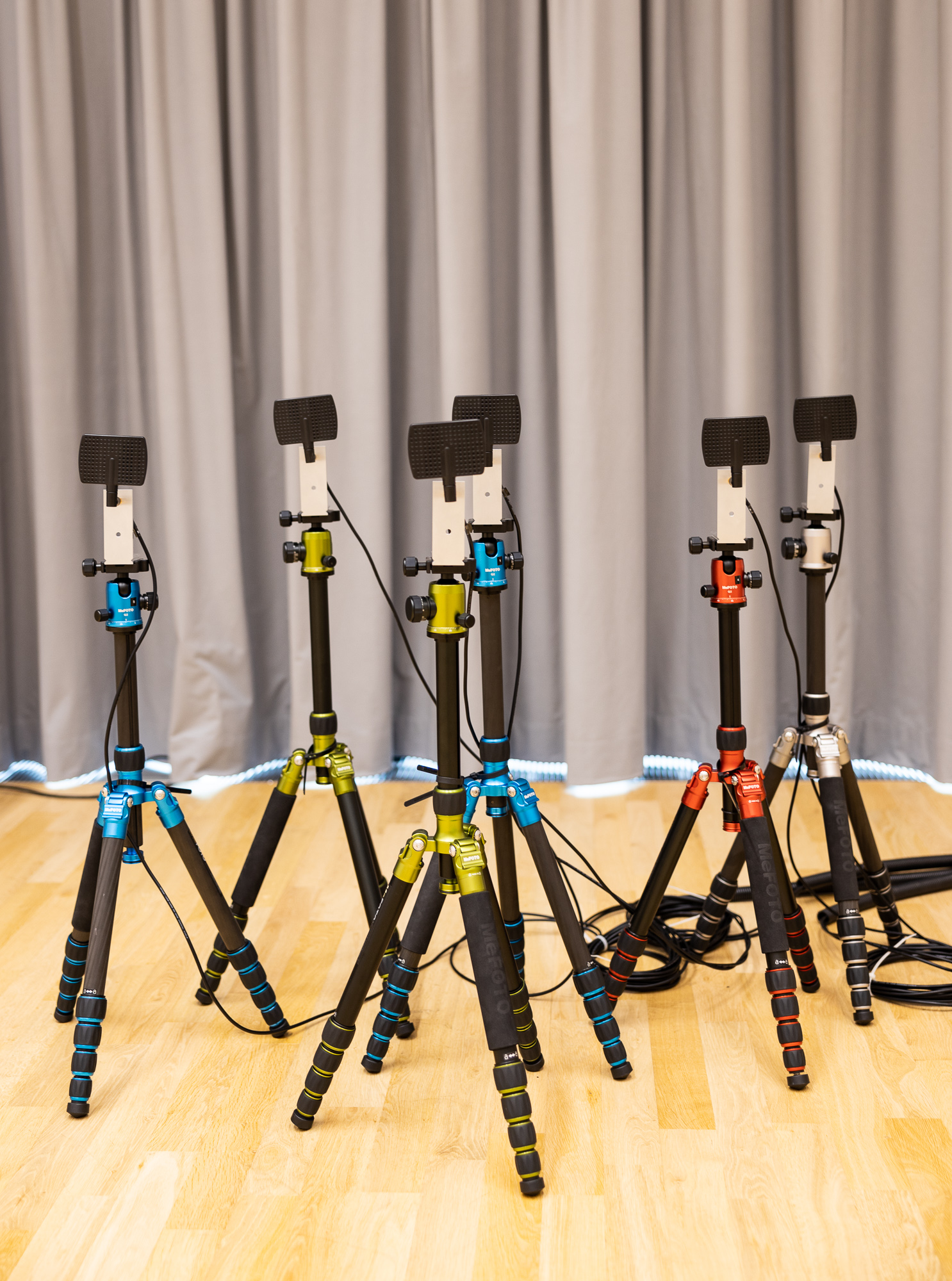}}}
        \caption{Users (\acl{Tx})}
        \label{fig:NOMAusers}
    \end{subfigure}
    \hfill
    \caption[NOMAsetup]{Real-time machine learning based multiuser detection setup.}
	\label{fig:NOMAsetup}
	%
    %
\end{figure}

\subsubsection{Users Antennas}
Each user antenna is installed on a tripod, allowing for conveniently adjusting location and height, relative to the receiving antenna array.
Every single antenna shown in \cref{fig:NOMAusers} represents a single transmitting user, sending its radio signal, and is connected to one of the eight transmitter ports on our \acs{SDR} module \cite{NAMC-SDR}.
According to the datasheet, the antenna gain is \SI[qualifier-mode=text]{5}{\deci\bel\isotropic} at \SI{2.5}{\giga\hertz} and up to \SI[qualifier-mode=text]{7}{\deci\bel\isotropic} at \SI{5.7}{\giga\hertz}, and the antenna polarization is vertical.

\subsubsection{Uniform Planar Array}
The \ac{UPA}, shown in \cref{fig:NOMAplanar}, is equipped with 32 cross-polarized patch antenna elements. These elements are arranged in 4 rows and 8 columns.
Each of 32 equidistant patch elements operates at a center frequency of \SI{2.442}{\giga\hertz}.

\subsubsection{Uniform Circular Array}
The sixteen physical \acs{Rx} antennas are arranged as a \ac{UCA}, as shown in \cref{fig:NOMAcircular}.
Uniform spacing is ensured by a ring retainer around the 16 antennas.
The antenna operates in the \SI{2.4}{\giga\hertz} and \SI{5}{\giga\hertz} \ac{WLAN} band with an omnidirectional radiation pattern and vertical polarisation.

%% file: sections/V_results.tex
\section{Performance Results}\label{sec:Res}

In this section, we present performance results for the experiments based on the setup presented in \cref{sec:SystemSetup}.
The objective is to demonstrate the real-time performance of the partially linear filtering algorithm presented in \cref{sec:apsm} in combination with the parallelization techniques and optimizations presented in \cref{sec:Implementation}.
In the following, we first show the performance of the partially linear filtering for various simulation settings and parameters.
We then compare the detection latencies of multiple platforms and show the acceleration in processing using techniques proposed in this study.

For the algorithm in \eqref{eqn:apsm}, unless noted otherwise, we use uniform Gaussian and linear weights, \ie, $w_\mathrm{L} = w_\mathrm{G} = 0.5$, a sliding window size of $W = 20$, and a Gaussian kernel variance of $\sigma^2 = 0.05$.
We use 16 antennas, and 6 users with uniform transmit power.
Each data point is an average of 100 transmissions, consisting of 685 training symbols and 3840 data symbols each.
The symbols consist of Gray-coded constellation points.
Modulated training and data symbols occupy up to 144 subcarriers on 5 \ac{OFDM} symbols, and 27 \ac{OFDM} symbols, respectively.
We use \SI{15}{\kilo\hertz} subcarrier spacing, which results in a total signal bandwidth of \SI{2.16}{\mega\hertz}.

\begin{figure}[t] 
    %
    %
    \centering
    \whencolumns{\def\figurewidth {0.70\linewidth}}{\def\figurewidth {0.97\linewidth}}
    \includegraphics[width=\figurewidth]{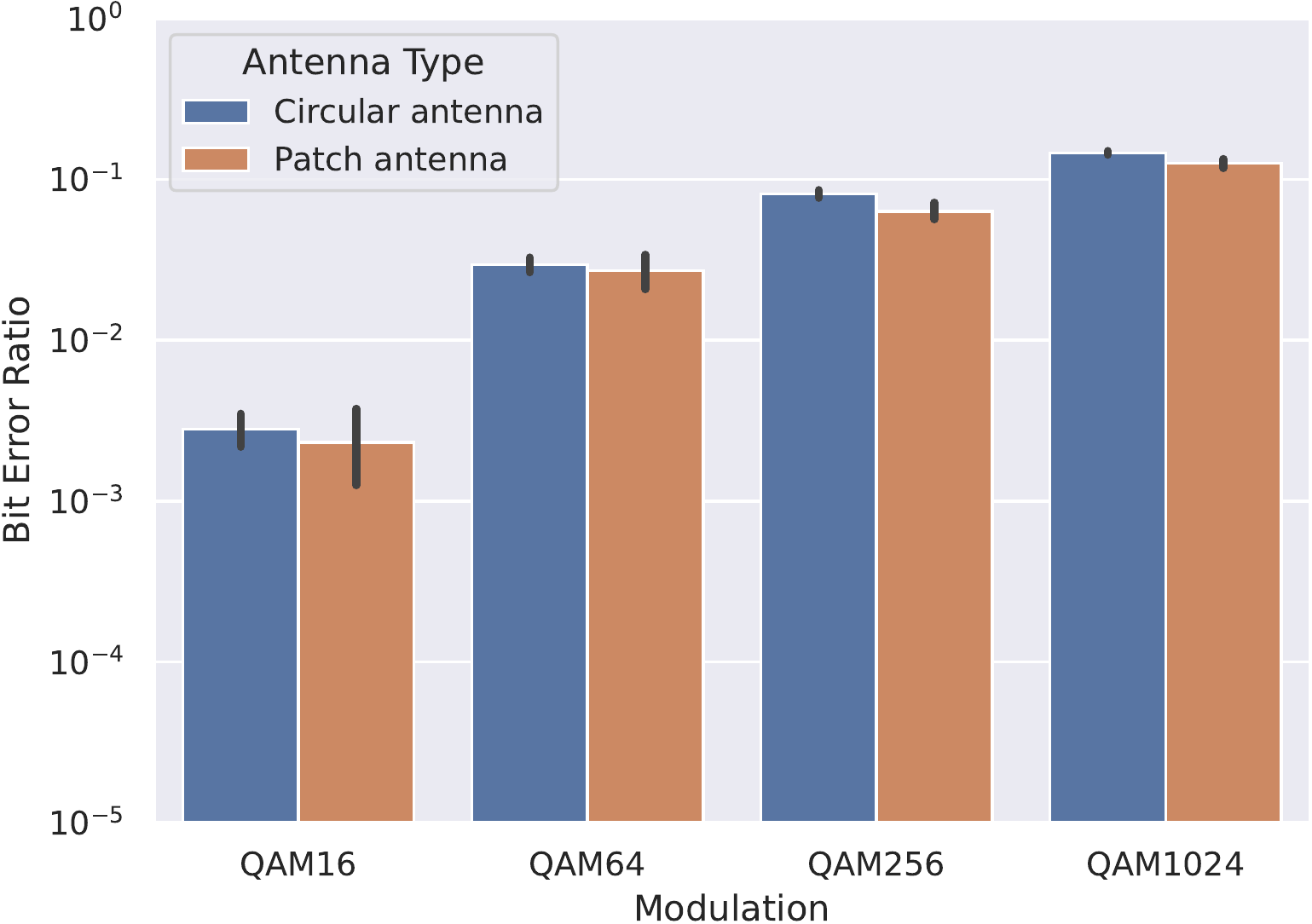}
    \caption{Average relative bit error rate for different modulation schemes. Black bars show the $\pm 1\mathrm{SD}$ confidence interval. BPSK and QPSK were omitted from the plot because the \ac{BER} is negligible.}
    \label{fig:plot_modulation}
    %
    %
\end{figure}

\Cref{fig:plot_modulation}, shows the \ac{BER} performance of standard modulation schemes.
All experiments were performed with the above-mentioned parameters. 
The error rate is negligible for \ac{BPSK} and \ac{QPSK}, while for higher-order modulations a significant number of bit errors occur.
As shown in \cref{fig:NOMAusers}, the angular separation of the users is very small, which  renders the task of separating them difficult.
Nevertheless, we observe that our algorithm can reach acceptable bit error rates even for high modulation schemes.
Note that in a practical system our algorithm would be followed by \ac{FEC}, which will correct the remaining bit errors.

\begin{figure}[t] 
    %
    %
    \centering
    \whencolumns{\def\figurewidth {0.70\linewidth}}{\def\figurewidth {0.97\linewidth}}
    \includegraphics[width=\figurewidth]{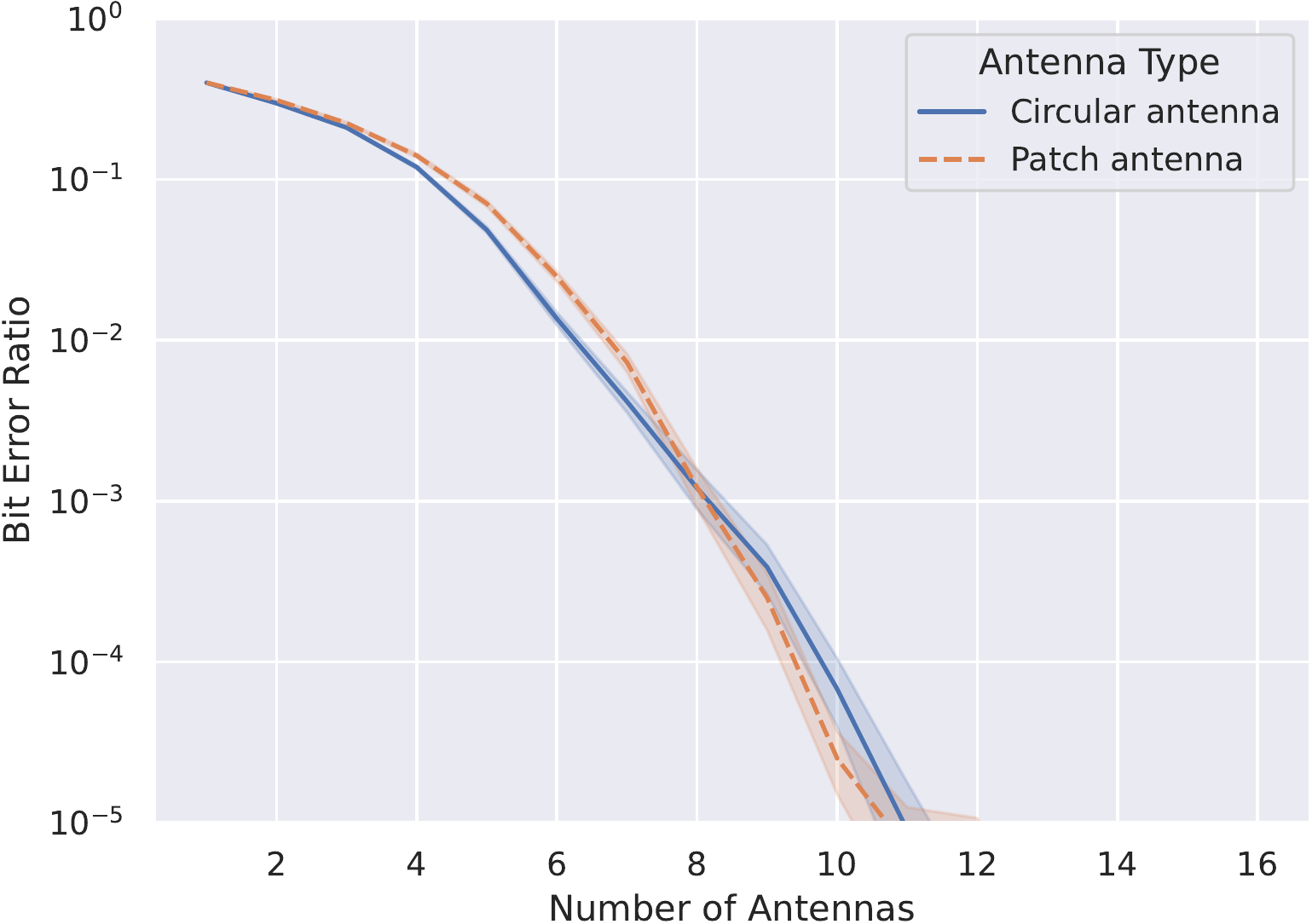}
    \caption{The average relative bit error rate for different numbers of antennas using \acs{QPSK}. Shaded regions show the $\pm 1\mathrm{SD}$ confidence interval.}
    \label{fig:plot_antennas}
    %
    %
\end{figure}

In \cref{fig:plot_antennas}, we demonstrate the \ac{BER} performance as a function of an increasing number of antennas, where the specified number of antennas was chosen randomly among the 16 antennas available in our system.
\comment{\footnote{\hl{We note that the \ac{BER} does not increase monotonically with the number of antennas because we did not physically change the antennas for each data point. Instead we selected a subset of antennas which was as close to equidistant as possible. However, this approach can lead to a suboptimal antenna configuration.}}.}
As mentioned in \cref{sec:apsm}, the algorithm does not assume any particular antenna array structure.
The measurements are executed for the two types of antenna arrays presented in \cref{sec:SystemSetup}.
We note that the results for both antenna types are very similar. 
We observe that the users become simpler to separate with an increasing number of antennas, and hence the bit error rate tends toward zero.
Even for a relatively small number of antennas as compared to the number of users, the performance is acceptable because the Gaussian kernels provide the capability to separate users that cannot be separated linearly.
\comment{%
Again, we can see that the performance degrades gracefully with the number of antennas, even though we did not change the configuration of the algorithm.
}%
\comment{%
We conclude that the presented algorithm is a very flexible solution, that can adapt to different scenarios without tuning the configuration.
In \cite{Awan_thesis} it was shown, that the proposed algorithm performs as well or better than \ac{SIC} while having significantly lower computational complexity.
Additionally, the proposed algorithm allows for fully parallel implementation, while \ac{SIC} has to be computed sequentially.
}%

\begin{table}[t] 
    \vspace*{+1.0\floatsep}
	\centering
        \whencolumns{\def\tablewidth {0.60\linewidth}}{\def\tablewidth {0.97\linewidth}}
        \begin{tabularx}{\tablewidth}{@{}*{1}{|r}*{3}{|C}|@{}}
    		\hhline{*{1}{~}*{3}{|-}|} 
    		\multicolumn{1}{r|}{}                             & {\cellcolor{NVLightGray}}RTX 2060~super
    		                                                  & {\cellcolor{NVLightGray}}RTX 2080~Ti
    		                                                  & {\cellcolor{NVLightGray}}Titan~V                                                 \\ \hhline{*{4}{|-}|}
    		{\cellcolor{NVLightGray}}Baseline Implementation  & \SI{7.82}{\milli\second} & \SI{3.676}{\milli\second} & \SI{2.23}{\milli\second}  \\ \hhline{*{4}{|-}|}
    		{\cellcolor{NVLightGray}}Multiple Vectors/Block   & \SI{7.76}{\milli\second} & \SI{2.180}{\milli\second} & \SI{2.37}{\milli\second}  \\ \hhline{*{4}{|-}|}
    		{\cellcolor{NVLightGray}}Shared Memory            & \SI{1.66}{\milli\second} & \SI{1.615}{\milli\second} & \SI{0.479}{\milli\second} \\ \hhline{*{4}{|-}|}
    		{\cellcolor{NVLightGray}}Balanced                 & \SI{1.25}{\milli\second} & \SI{0.906}{\milli\second} & \SI{0.403}{\milli\second} \\ \hhline{*{4}{|-}|}
        \end{tabularx}
	\caption{Summary of the performance improvements of the detection \acs{CUDA} kernel for various optimization steps as described in \cref{sec:Optimization}.}
    \label{tab:optimization_results}
	%
	%
\end{table}

In \cref{tab:optimization_results}, we present the detection latencies of the different implementations presented in \cref{sec:Optimization} on different \acp{GPU}.
The \acp{GPU} represent two different device families: The RTX~2060~super and RTX~2080~Ti are consumer-grade boards with \ac{GDDR} memory, while the Titan~V is a data-center class board equipped with \ac{HBM} and with more computing resources.
All used \ac{GPU} cards execute the same code (neither device-specific optimizations nor tuning was carried out).
The table does not show the time elapsed during training, which amounts to roughly \SI{100}{\milli\second}.
As mentioned in Remark~\ref{sec:training_tracking_detection}, the relatively long training time is not of concern because retraining is, in principle, only required if the environment changes abruptly. 
\comment{%
This is because performance is mostly unmodified by the different versions and mainly dependent on the number of iterations performed by the kernel, which is dependent on the number of samples present in the training set as well as the window size and the sweep of the window.
}%
For comparison, a \acs{MATLAB} implementation of our algorithm takes more than \SI{40}{\second} to execute on an i7-6700T CPU with simple parallelization (up to 8 Threads on 4 Cores) from the native Parallel Toolbox. The comparison with signal processing in \acs{MATLAB} may not be fair, because we did not optimize the \acs{MATLAB} code for speed.
Nevertheless, one may conclude that the optimized \ac{CUDA} implementation is several orders of magnitude faster than a native \ac{CPU}-based implementation.
\comment{%
Moreover,
However, a naive implementation that does not exploit the structure of the particular algorithm at hand may not achieve the desired results.
In contrast, our optimized solution (see \cref{sec:Implementation} for details) achieves the desired goal of an extremely low latency suitable for \ac{ULL} systems.
}%


%% file: sections/VI_conclusions.tex


\section{Conclusions}\label{sec:Con}
Our proof-of-concept provides a practical and fast implementation of recently proposed partially linear adaptive filtering for \acl{MU} detection on \ac{GPU}-accelerated platforms.
We exploit the parallelism intrinsic to the mathematical formulation and distribute the computations in an ``optimal'' way across the targeted \ac{GPU}.
The techniques developed for hiding latency and accelerating memory access are key to reaching real-time performance.
As a result, we can perform \acl{MU} detection with a detection latency of below \SI{1}{millisecond} with our \ac{COTS} laboratory setup.
In future work, similar techniques will be applied to drastically reduce the total amount of processing time spent on training.
Moreover, channel coding will be included in the future. Finally, we note that \ac{APSM} shares many operations with similar projection-based algorithms in Hilbert spaces which have seen many applications in signal processing and machine learning.
Therefore, our acceleration techniques may be generalized to such projection-based algorithms and algorithms based on \aclp{RKHS}.
The developed and used \ac{CUDA} \ac{APSM} library code is publicly available on the GitHub server page of the Fraunhofer HHI \cite{CUDA_libapsm}.


%% file: sections/acknowledgments.tex

\section*{Acknowledgments}
\addcontentsline{toc}{section}{Acknowledgments}

This work has been partially funded by the German Federal Ministry of Education and Research (BMBF, Germany) in the project Open Testbed Berlin - 5G and Beyond (OTB-5G+) under Grant 16KIS0980 and supported as part of the 6G Research and Innovation Cluster 6G-RIC under Grant 16KISK020K.


%% file: sections/references.tex




\bibliographystyle{IEEEtran}

\bibliography{%
    bib/noma_demonstrator_book,
    bib/noma_demonstrator_thesis,
    bib/noma_demonstrator_paper,
    bib/noma_demonstrator_techrep,
    bib/noma_demonstrator_url,
    IEEEabrv
}


%% file: 2022_noma_cuda_demonstrator_arXiv.bbl
\begin{thebibliography}{10}
\providecommand{\url}[1]{#1}
\csname url@samestyle\endcsname
\providecommand{\newblock}{\relax}
\providecommand{\bibinfo}[2]{#2}
\providecommand{\BIBentrySTDinterwordspacing}{\spaceskip=0pt\relax}
\providecommand{\BIBentryALTinterwordstretchfactor}{4}
\providecommand{\BIBentryALTinterwordspacing}{\spaceskip=\fontdimen2\font plus
\BIBentryALTinterwordstretchfactor\fontdimen3\font minus
  \fontdimen4\font\relax}
\providecommand{\BIBforeignlanguage}[2]{{%
\expandafter\ifx\csname l@#1\endcsname\relax
\typeout{** WARNING: IEEEtran.bst: No hyphenation pattern has been}%
\typeout{** loaded for the language `#1'. Using the pattern for}%
\typeout{** the default language instead.}%
\else
\language=\csname l@#1\endcsname
\fi
#2}}
\providecommand{\BIBdecl}{\relax}
\BIBdecl

\bibitem{Shin2017}
W.~Shin, M.~Vaezi, B.~Lee, D.~J. Love, J.~Lee, and H.~V. Poor, ``Non-orthogonal
  multiple access in multi-cell networks: Theory, performance, and practical
  challenges,'' \emph{IEEE Communications Magazine}, vol.~55, no.~10, pp.
  176--183, October 2017.

\bibitem{Wang2016}
Y.~Wang, B.~Ren, S.~Sun, S.~Kang, and X.~Yue, ``Analysis of non-orthogonal
  multiple access for 5{G},'' \emph{China Communications}, vol.~13, no.
  Supplement2, pp. 52--66, N 2016.

\bibitem{Ding2017}
Z.~Ding, X.~Lei, G.~K. Karagiannidis, R.~Schober, J.~Yuan, and V.~K. Bhargava,
  ``A survey on non-orthogonal multiple access for 5{G} networks: Research
  challenges and future trends,'' \emph{CoRR}, vol. abs/1706.05347, 2017.

\bibitem{Tabassum2016}
\BIBentryALTinterwordspacing
H.~Tabassum, M.~S. Ali, E.~Hossain, M.~J. Hossain, and D.~I. Kim,
  ``Non-orthogonal multiple access {(NOMA)} in cellular uplink and downlink:
  Challenges and enabling techniques,'' \emph{CoRR}, vol. abs/1608.05783, 2016.
  [Online]. Available: \url{http://arxiv.org/abs/1608.05783}
\BIBentrySTDinterwordspacing

\bibitem{HIGUCHI2015}
K.~Higuchi and A.~Benjebbour, ``Non-orthogonal multiple access ({NOMA}) with
  successive interference cancellation for future radio access,'' \emph{IEICE
  Transactions on Communications}, vol. E98.B, no.~3, pp. 403--414, 2015.

\bibitem{Xin2016}
X.~Su, H.~Yu, W.~Kim, C.~Choi, and D.~Choi,
  ``\BIBforeignlanguage{eng}{Interference cancellation for non-orthogonal
  multiple access used in future wireless mobile networks},''
  \emph{\BIBforeignlanguage{eng}{EURASIP journal on wireless communications and
  networking}}, vol. 2016, no.~1, pp. 1--12, 2016.

\bibitem{Islam2017}
S.~M.~R. Islam, N.~Avazov, O.~A. Dobre, and K.-s. Kwak, ``Power-domain
  non-orthogonal multiple access ({NOMA}) in {5G} systems: Potentials and
  challenges,'' \emph{IEEE Communications Surveys Tutorials}, vol.~19, no.~2,
  pp. 721--742, 2017.

\bibitem{9148869}
M.~{Mehlhose}, D.~A. {Awan}, R.~L.~G. {Cavalcante}, M.~{Kurras}, and
  S.~{Stanczak}, ``Machine learning-based adaptive receive filtering:
  Proof-of-concept on an {SDR} platform,'' in \emph{ICC 2020 - 2020 IEEE
  International Conference on Communications (ICC)}, 2020, pp. 1--5.

\bibitem{8422449}
D.~A. {Awan}, R.~L.~G. {Cavalcante}, M.~{Yukawa}, and S.~{Stanczak},
  ``Detection for {5G}-{NOMA}: An online adaptive machine learning approach,''
  in \emph{2018 IEEE International Conference on Communications (ICC)}, May
  2018, pp. 1--6.

\bibitem{Awan_thesis}
\BIBentryALTinterwordspacing
D.~A. Awan, ``Robust learning in wireless networks : efficacy of models and
  prior knowledge in learning from small sample sets,'' Doctoral Thesis,
  Technische Universität Berlin, Berlin, 2021. [Online]. Available:
  \url{http://dx.doi.org/10.14279/depositonce-11266}
\BIBentrySTDinterwordspacing

\bibitem{Bjornson2017}
E.~Bjornson, J.~Hoydis, and L.~Sanguinetti, \emph{Massive MIMO Networks:
  Spectral, Energy, and Hardware Efficiency}.\hskip 1em plus 0.5em minus
  0.4em\relax Now Foundations and Trends, 2017.

\bibitem{Yamada}
\BIBentryALTinterwordspacing
I.~Yamada and N.~Ogura, ``Adaptive projected subgradient method for asymptotic
  minimization of sequence of nonnegative convex functions,'' \emph{Numerical
  Functional Analysis and Optimization}, vol.~25, no. 7-8, pp. 593--617, 2005.
  [Online]. Available: \url{https://doi.org/10.1081/NFA-200045806}
\BIBentrySTDinterwordspacing

\bibitem{Combettes93}
P.~Combettes, ``The foundations of set theoretic estimation,''
  \emph{Proceedings of the IEEE}, vol.~81, no.~2, pp. 182--208, 1993.

\bibitem{Stark1998}
H.~Stark, Y.~Yang, and Y.~Yang, \emph{Vector Space Projections: A Numerical
  Approach to Signal and Image Processing, Neural Nets, and Optics}.\hskip 1em
  plus 0.5em minus 0.4em\relax New York, NY, USA: John Wiley \& Sons, Inc.,
  1998.

\bibitem{nvidia_web_gpu}
\BIBentryALTinterwordspacing
{NVIDIA} {Aerial}: Build and deploy {GPU}-accelerated 5{G} virtual {Radio}
  {Access} {Networks} (v{RAN}). [Online]. Available:
  \url{https://developer.nvidia.com/aerial-sdk}
\BIBentrySTDinterwordspacing

\bibitem{Yukawa2015}
M.~Yukawa, ``Adaptive learning in cartesian product of reproducing kernel
  {H}ilbert spaces,'' \emph{IEEE Transactions on Signal Processing}, vol.~63,
  no.~22, pp. 6037--6048, Nov 2015.

\bibitem{Yamada2005}
I.~Yamada and N.~Ogura, ``Adaptive projected subgradient method for asymptotic
  minimization of sequence of nonnegative convex functions,'' \emph{Numerical
  Functional Analysis and Optimization}, vol.~25, no. 7-8, pp. 593--617, 2005.

\bibitem{Slavakis2009}
K.~Slavakis, S.~Theodoridis, and I.~Yamada, ``Adaptive constrained learning in
  reproducing kernel {H}ilbert spaces: The robust beamforming case,''
  \emph{IEEE Transactions on Signal Processing}, vol.~57, no.~12, pp.
  4744--4764, Dec 2009.

\bibitem{Theodoridis2011}
S.~Theodoridis, K.~Slavakis, and I.~Yamada, ``Adaptive learning in a world of
  projections,'' \emph{IEEE Signal Processing Magazine}, vol.~28, no.~1, pp.
  97--123, Jan 2011.

\bibitem{Du2017}
J.~Du and R.~A. Valenzuela, ``How much spectrum is too much in millimeter wave
  wireless access,'' \emph{IEEE Journal on Selected Areas in Communications},
  vol.~35, no.~7, pp. 1444--1458, July 2017.

\bibitem{CUDA_Programming_Guide}
\BIBentryALTinterwordspacing
{CUDA} {Programming} {Guide}. [Online]. Available:
  \url{https://docs.nvidia.com/cuda/cuda-c-programming-guide/index.html}
\BIBentrySTDinterwordspacing

\bibitem{NAMC-SDR}
\BIBentryALTinterwordspacing
{N.A.T. GmbH} {NAMC-SDR}. [Online]. Available:
  \url{http://www.nateurope.com/products/NAMC-SDR.html}
\BIBentrySTDinterwordspacing

\bibitem{CUDA_libapsm}
\BIBentryALTinterwordspacing
{CUDA} {APSM} {library}. [Online]. Available:
  \url{https://github.com/fraunhoferhhi/libapsm}
\BIBentrySTDinterwordspacing

\end{thebibliography}
